\documentclass[journal,twocolumn]{IEEEtran}
\usepackage[utf8]{inputenc}
\usepackage{amsmath,mathtools}
\usepackage{bbold}
\usepackage{xcolor}
\usepackage{bbm}
\usepackage{graphicx}
\usepackage{physics}
\usepackage{cite}
\usepackage{amssymb}
\usepackage{epstopdf}
\usepackage{amsthm}
\usepackage{subfigure}
\usepackage{cleveref}
\usepackage[linesnumbered,ruled,vlined]{algorithm2e}
\usepackage{algorithmic, float,setspace}
\newtheorem{lemma}{Lemma}
\newtheorem{theorem}{Theorem}
\newtheorem{remark}{Remark}

\newtheorem{corollary}{Corollary}

\usepackage{tcolorbox} 
\allowdisplaybreaks

\DeclareMathOperator*{\argmax}{arg\,max}

\newtcolorbox{mybox}{colback=red!0!white,colframe=red!0!black} 
\newtcolorbox{myboxtitle}[1]{colback=yellow!5!white,colframe=orange!90!yellow,fonttitle=\bfseries,title=#1} 

\SetKwInput{KwResult}{Result}                
\SetKwInput{KwOutput}{output}              
\SetKwInput{KwInput}{input}              
\SetKwInput{KwInitialize}{initialize}
\SetKwInput{KwIntended}{intended entity}
\SetKwRepeat{Repeat}{repeat}{until}
\begin{document}
\bstctlcite{IEEEexample:BSTcontrol}

\title{Semantics-Native Communication via Contextual Reasoning}

\author{Hyowoon~Seo,~\IEEEmembership{Member,~IEEE},
Jihong~Park,~\IEEEmembership{Senior Member,~IEEE},
Mehdi~Bennis,~\IEEEmembership{Fellow,~IEEE},
and M\'erouane~Debbah,~\IEEEmembership{Fellow,~IEEE}
\thanks{H. Seo is with the Department of Electronics and Communications Engineering, Kwangwoon University, Seoul 01897, Korea (e-mail: hyowoonseo@kw.ac.kr).}
\thanks{J. Park is with the School of Information Technology, Deakin University, Geelong, VIC 3220, Australia (email: jihong.park@deakin.edu.au).}
\thanks{M. Bennis is with the Centre for Wireless Communications, University of Oulu, Oulu 90014, Finland (e-mail: mehdi.bennis@oulu.fi).}
\thanks{M\'erouane Debbah is with Technology Innovation Institute, Abu Dhabi, UAE, and also with the Centrale Supelec, University Paris-Saclay, 91192 Gifsur-Yvette, France (e-mail: merouane.debbah@tii.ae).}
}

\maketitle

\vspace{-0.2in}

\begin{abstract}
Recently, machine learning (ML) has shown its effectiveness in improving communication efficiency by reinstating the semantics of bits. To understand its underlying principles, we propose a novel stochastic model of semantic communication, dubbed \emph{semantics-native communication (SNC)}. Inspired from human communication, we consider a point-to-point SNC scenario where a speaker has an intention of referring to an entity, extracts its semantic concepts, and maps these concepts into symbols that are communicated to a target listener through the traditional Shannon communication channel. Next, we recall rational humans who can communicate more efficiently by reasoning about the others' contexts before communication, referred to as contextual reasoning. This motivates us to propose a novel SNC framework harnessing agent states as side information in a way that the speaker locally and iteratively communicates with a virtual agent having the listener's state, and vice versa. Theoretically, we prove the convergence of contextual reasoning, at which it minimizes the bit length while guaranteeing a target reliability. Simulation results corroborate that contextual reasoning based SNC can significantly reduce bit lengths, and be a robust solution to imperfect agent states by quantizing the entity-concept-symbol mapping before contextual reasoning.

\end{abstract}



\section{Introduction}

Spurred by the advances in machine learning (ML), communication systems become more intelligent, and the data transmission accuracy is expect not to solely determine the communication quality in the upcoming sixth-generation (6G) communication systems. Instead, akin to human communications, it becomes important to deal with the problem of maximizing the \emph{effectiveness} of communications by conveying the \emph{semantics} (or meanings) of data \cite{Shannon1964}. Shannon himself has identified these two issues together as the \emph{semantic-effectiveness problem}, departing from the traditional \emph{technical problem} of accurately transferring and reconstructing data in the current communication systems. 
Recently, there have been several studies on communication for addressing this semantic-effectiveness problem \cite{Popovski2019,Kountouris2020,Maatouk2020,Yun2021,Kalfa2021}, referred to as semantics-empowered communication, goal-oriented communication, or more generally \emph{semantic communication (SC)}. The main focus of these works is to deliver meaningful messages that are effective in a given task as intended. 

However, the existing SC studies commonly postulate that each listener's decoding process is just a mirror image of its speaker's encoding process. This is valid in traditional communication systems where common encoder-decoder architectures are pre-programmed and shared across all communicating agents. In stark contrast, the encoders and decoders in ML based SC frameworks are expected to be separately trained using their local data samples. This leads to the heterogeneous context of individual speakers and listeners, which are unmodeled in the existing works. In fact, ML based solutions can cope with this issue by learning such heterogeneous contexts through iterative communications \cite{Foerster2016,Lazaridou2017,Lazaridou2020}. Notwithstanding, ML processes are in a black box, not allowing us to intuitively understand and improve SC operations.

\begin{figure*}
\centering
\subfigure[System 1 SNC.]{\includegraphics[width=0.45\textwidth]{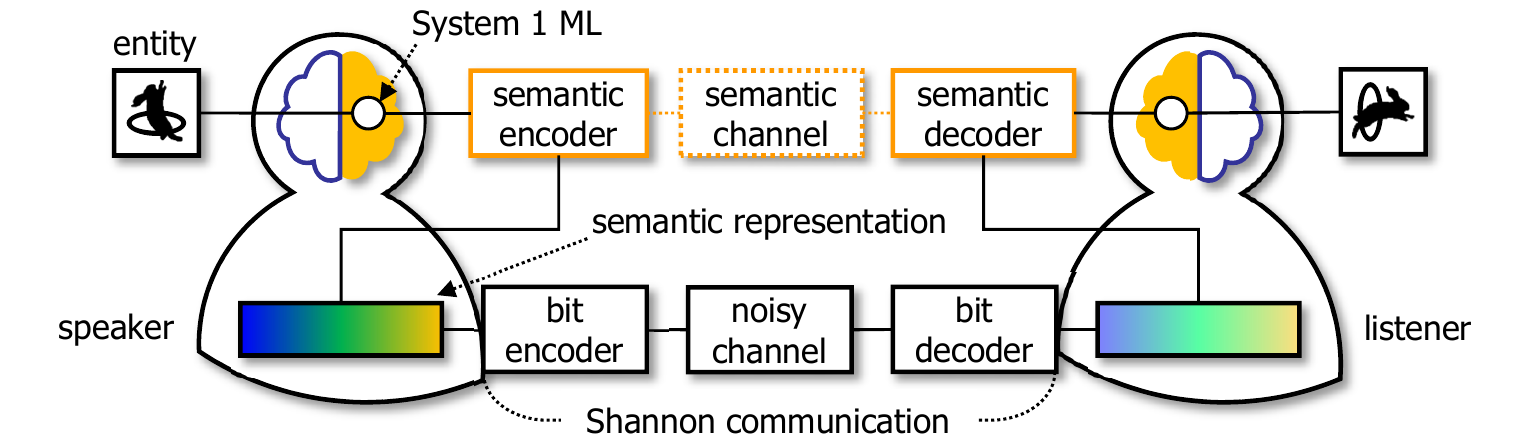}\label{fig:system1}}\\
\subfigure[System 2 SNC.]{\includegraphics[width=0.45\textwidth]{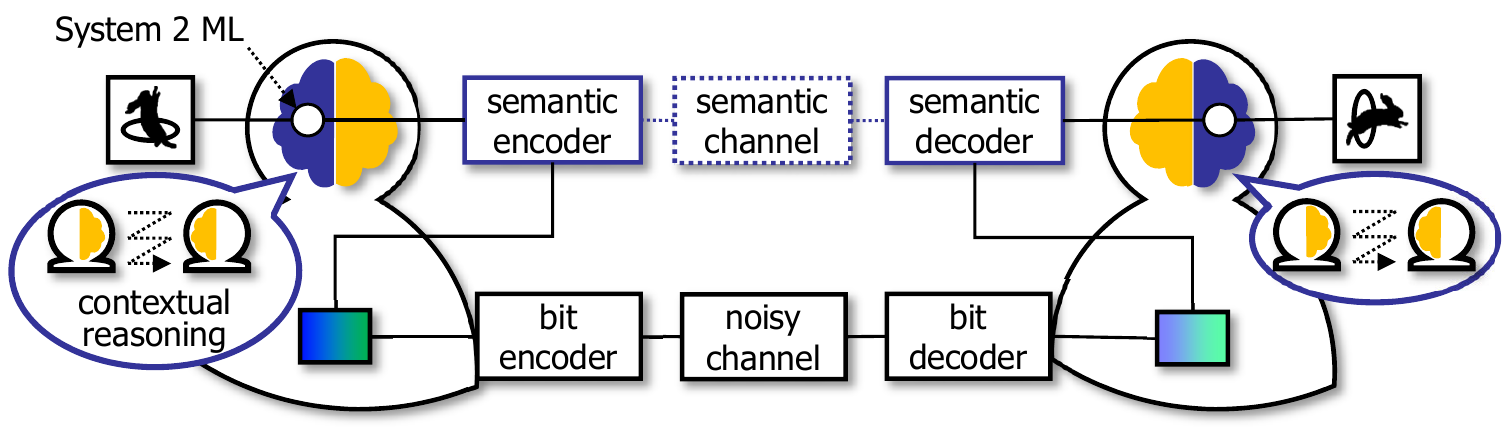}\label{fig:system2}}
\caption{A schematic illustration of (a) System 1 stochastic semantic communication (SNC), whose semantic encoder conceptualizes and symbolizes a communicating action and semantic decoder vice versa, and (b) System 2 SNC model, whose semantic encoder and decoder are instilled with contextual reasoning.}
	\label{Fig_Shannon_semantic}
	\vspace{-15pt}
\end{figure*}

To fill this void, the overarching goal of this article is to open up the black-box of SC between intelligent agents (e.g., across ML-driven machines or between machines and humans) and examine the impact of communication context thereon by building a mathematical model. To this end, inspired by linguistics and information theory, we propose a novel stochastic \emph{semantics-native communication (SNC)} model where a speaker maps an entity of interest (e.g., abstract idea, physical phenomena and objects) into semantic concepts and then into symbols, resulting in a \emph{semantic representation (SR)}. At a speaker, these \emph{semantic encoding} operations are introduced before the classical source/channel encoding in Shannon's theory, as illustrated in Fig.~\ref{Fig_Shannon_semantic}. Likewise, a listener performs \emph{semantic decoding} operations before the classical source/channel decoding, thereby reproducing the intended action from the received SR. The inputs and outputs of such semantic coding and the classical source/channel coding are connected, allowing us to analyze their in-depth operations and making inroads towards advancing the principles of SC.

Treating the aforementioned SNC model as \emph{System~1 SNC}, we additionally put forward \emph{System~2 SNC}, spurred by the recent paradigm shift from System~1 ML to System~2 ML \cite{Bengio2019}. According to \cite{Bengio2019} itself inspired by Daniel Kahneman's book \emph{Thinking, Fast and Slow} \cite{Kahneman2011}, System 1 ML is tantamount to fast and unconscious pattern recognition as in the current deep learning, in which semantic coding in System 1 SNC (hereafter referred to as \emph{System 1 semantic coding}), falls. In contrast, System 2 ML is about slow and logical metacognition, which enables reasoning, planning and handling exceptions. Inspired by this, System 2 SNC incorporates \emph{System 2 semantic coding} to infuse \emph{contextual reasoning} into System 1 SNC, such that each agent runs internal simulations by locally and iteratively reasoning about the communication context of its communication counterpart before every communication round, analogous to `\emph{I} think of \emph{You} thinking of \emph{Me} thinking of \emph{You} and so on.' Such contextual reasoning \cite{Scholkopf2021,Zaslavsky2020,Wang2020,Kang2020} gives rise to more efficient way of communication between agents. This is specialized for its unique ways of semantic coding induced by different tasks and other communication contexts. As a result, compared to System 1 SNC that conveys all the semantics associated with the action of interest, System 2 SNC significantly reduces the communication cost by sending only the most effective semantics to its communication counterpart.

As Fig.~\ref{Fig_Shannon_semantic} illustrates, System 1 semantic coding is performed outside of the classical source and channel coding, hereafter referred to as \emph{Shannon coding}. Hence, any bit errors due to Shannon coding may result in semantic errors, emphasizing the important role of Shannon coding. While inheriting this property, System 2 semantic coding additionally aims to address the semantic errors due to the contextual difference between the speaker and its listener. Such type of contextual semantic errors cannot be corrected by simply adding more bits as opposed to Shannon coding and System 1 semantic coding; for instance, via forward error correction for channel coding and with less compression for source coding and System 1 semantic coding. The required solution for these contextual semantic errors is in contrast to Shannon coding and System 1 semantic coding which commonly postulate that decoding is a mirror image of encoding, highlighting the importance and uniqueness of System~2 semantic coding.

\subsection{Related Works}
\textbf{Semantic Communication.}\quad 
The aforementioned semantic and effectiveness problems have been identified just a year after the inception of the Shannon communication model \cite{Shannon1964}. Spurred by ML, the problem has recently been revisited through the lens of the \textit{semantics-empowered communication} framework that is broadly categorized into three subdirections. The first direction is to filter out semantically unimportant information (e.g., outdated control commands or texture details for shape detection)
before transmission \cite{Popovski2019,Kountouris2020,Maatouk2020,Yun2021}. As opposed to the classical compression in source coding, meaningfulness is determined not solely by raw data characteristics \cite{Kalfa2021}, but jointly by the listener's goal-oriented metrics, e.g., age-of-incorrect information (AoII) \cite{Kountouris2020,Maatouk2020} and control stability \cite{Popovski2019,Weihang2019}, as well as by the mutual context between agents, e.g., attention-based similarity \cite{Yun2021}. The second direction lies in transforming the modality of raw data while maintaining the same meaning. This includes the image-to-text embedding via Transformer \cite{Xie2021a,Xie2021b} and the conversion from system states into the state evolution law using the Koopman operator \cite{Girgis2021,Girgis2022}. Lastly, one can build a knowledge base to exploit the knowledge as side information or to leverage dependencies in target data. In doing so, communication payload sizes can be reduced\cite{choi2022unified}, which was demonstrated in speech-based communication, video streaming, and holographic communication applications \cite{Strinati2021,Shi2021}.

Meanwhile, SC has been taken into consideration in the context of the \textit{emergent communication} framework to create new semantic vocabularies and syntax for ML-driven agents such that their communication becomes effective in their downstream tasks. Technically, these methods are based mostly on multi-agent reinforcement learning (MARL), where interactions among agents induce the emergence of semantic communication. One prominent direction is the differentiable inter-agent learning (DIAL) framework under a continuous communication channel \cite{Foerster2016,Singh2019,Kim2019,Lee2022}, where different agents’ ML models are concatenated into a single model that is trained using the backpropagation algorithm. Another direction is reinforced inter-agent learning (RIAL) type methods under a discrete communication channel \cite{Foerster2016,Lowe2017,Lazaridou2017} where the standard back-propagation algorithm is not applicable.


In a nutshell, the applications of existing semantics-empowered communication frameworks are often restricted to specific data domains (e.g., images and natural languages) and/or environments (e.g., control systems). On the other hand, existing emergent communication frameworks are relatively flexible and applicable to a broader range of scenarios, but analysis on their ML-based internal operations including algorithmic convergence is lacking, calling for a principled approach. {Motivated by these limitations, in this article, we develop a novel stochastic model of SC, based on the premise that training an ML model is equivalent to finding a conditional distribution between the input data and their desired outputs. Such a stochastic model is compatible with the standard Shannon communication model, allowing us to analyze its end-to-end communication operations. Inspired by the two levels of human cognition, we treat the baseline SNC model as System 1 SNC, and additionally put forward its System 2 SNC. System 2 SNC reflects the heterogeneous contexts of agents before actual communication via locally and internally communicating with their virtual communication counterparts. Such contextual reasoning  obtains an emergent language that significantly improve the communication efficiency and effectiveness of SNC.}

\textbf{Contextual Reasoning.}\quad According to \cite{Bengio2019}, System 1 ML -- which is commonly fast, intuitive and unconscious -- is about recognizing patterns and statistical correlations in raw data space. In contrast, System 2 ML -- which is slow, logical and conscious -- is about finding the underlying causation of the perceived correlations while reasoning on the data’s SRs \cite{Scholkopf2021}. The next paradigm shift from System~1~ML to System 2 ML based communication systems that harnesses logical reasoning for communication is upon us. 
Among various types of reasoning \cite{Goyal2020,Scholkopf2021}, in this work we mainly focus on incorporating principles of \emph{contextual reasoning} (or pragmatic reasoning) into SNC. Contextual reasoning refers to the humans' ability to reason about the hidden meaning beyond communication, such as linguistic ambiguity and intention of the interlocutor, based on the local context of the communication and social-interactions \cite{Grice1975,Bell1995,Bell1999,Bell2001}. A well-known computational approach to contextual reasoning is the rational speech act (RSA) framework \cite{Frank2012, Goodman2013, Goodman2016, Frank2016}, which formulates the speaker and listener as stochastic models and simulates contextual reasoning based human communication. Recently, the connection between the RSA model and optimal transport was explored in \cite{Wang2020} and investigated from the information-theoretic rate-distortion point of view in \cite{Zaslavsky2020}. While these computational approaches are interesting, as `RSA' literally implies, the model focuses mostly on the speaker, ignoring the listener and its interactions with the speaker. In contrast, we primarily focus on contextual reasoning of both speaker,  listener and  their interactions, and  build a computational framework of such reasoning.

\subsection{Contributions and Organization}
The major contributions of this work are summarized as follows.
\begin{itemize}
    \item We propose a novel stochastic model of point-to-point SC, named System 1 SNC, and derive a closed-form expression of the expected bit-length of the SR (see \textbf{Theorem 1}).
    
    \item Next, we propose System 2 SNC by infusing contextual reasoning into System 1 SNC, while formulating the contextual reasoning as an optimization problem with respect to agents' local contexts. We develop a locally iterative solution to the contextual reasoning, and prove that both agents' states converge to a common communication context (see \textbf{Theorem 2}).
    
    \item Finally, leveraging the proposed stochastic model, we show that the reliability of System 2 SNC increases with the number of meaningful concepts (see \textbf{Theorem~3}), and derive the expected SR bit length under System 2 SNC (see \textbf{Corollary~3}).
\end{itemize}

The rest of this article is organized as follows. The stochastic models of System 1 and System 2 SNC are formalized and analyzed in Sections \ref{sec:semantic_communication} and \ref{sec:RHSC}, respectively. The effectiveness of System 2 SNC is corroborated in Section \ref{sec:experiments} via experimental results based on the proposed stochastic model. Finally, this article is concluded in Section \ref{sec:discussion}.

\section{Semantics-Native Communication Framework}\label{sec:semantic_communication}
{Consider a world wherein a pair of agents carry out tasks that require communication (or coordination).\footnote{{An illustrative example of such a communication task is an image referential game, in which a speaker talk about one among the given set of images and the listener guesses what the image is.}}} To communicate, one agent becomes a \emph{speaker} and the other becomes a \emph{listener}, and the goal is for the speaker to induce the listener to take a specific action. The effectiveness of the communication is determined by whether the listener's action is well-aligned with the speaker's intention. The communication process of the speaker and the listener is stochastically modeled and will be hereafter referred to as a \emph{System~1~SNC} model.

\begin{figure}\centering
	\includegraphics[width=0.4\textwidth]{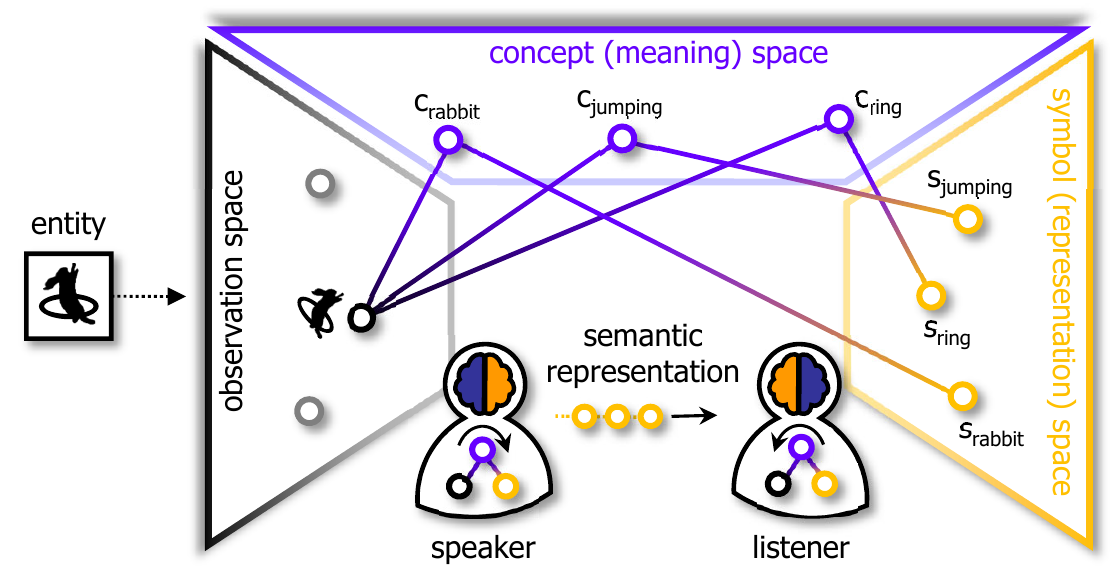}
	\caption{An illustration of the multi-triangular semantic coding model in SNC, inspired by the Ogden \& Richards' semantic triangle \cite{Ogden1923}.}
	\label{Fig_triangles}
	\vspace{-15pt}
\end{figure}

The modeling of System 1 SNC is inspired from the Ogden \& Richards' \emph{the triangle of meanings} \cite{Ogden1923} in human cognition and communication. As visualized in Fig. \ref{Fig_triangles}, the vertices of this semantic triangle connect the three spaces of the actions,\footnote{There can be any other entities that the agents refer to, such as abstract ideas, physical phenomena, etc.} concepts (or semantics), and symbols (or representation). The directed edge from an action to a concept is called \emph{conceptualization}, and the edge from the concept to a symbol is called \emph{symbolization}, while their opposite directions imply deconceptualization and desymbolization, respectively.

{Borrowing this model into System 1 SNC, we define a \emph{concept} as a unit of an agent's interpretation that partly or fully describes the intrinsic properties of an action in the semantic domain, and a \emph{symbol} as a concept's representation in a communicable form.\footnote{For example, an image referential game, concepts and symbols can be seen as extracted features from an image and their representation, respectively.}}  Let $\mathcal{A}$ be the finite set of \emph{actions} that a listener can take, and assume that $\mathcal{A}$ is commonly known by both the speaker and listener. Consider that each action connotes one or multiple concepts in a finite set $\mathcal{C}$ of the entire world concepts, which is known to all agents. Consequently, System 1 SNC can be summarized by a multi-triangular model with a shared intended action as illustrated in Fig. \ref{Fig_triangles}. Here, \emph{semantic encoding} encompasses conceptualization and symbolization, i.e., action-to-symbol model, while \emph{semantic decoding} incorporates deconceptualization and desymbolization, i.e., symbol-to-action model, as detailed next.



\subsection{System 1 Semantic Coding Model}\label{subsec:semantic_coding}
\subsubsection{Action-Concept Relevance}
Once a speaker selects an intended action $a \in \mathcal{A}$, it extracts multiple relevant concepts from $a$. These relevant concepts are not always identical to those of the other agents carrying out different downstream tasks. {To reflect this, we suppose that each agent is parameterized by a \emph{task index} $t\in \mathcal{T}$ and assume the number of tasks in the world is finite. It is worth mentioning that the agents that deal with the same tasks are parameterized by the same task index.}
Let $X_c \in \{\mathtt{TRUE},\mathtt{FALSE}\}$ be a binary random variable that indicates whether a concept $c\in \mathcal{C}$ is relevant ($X_c = \mathtt{TRUE}$) or not ($X_c = \mathtt{FALSE}$) to a certain action, out of the finite set $\mathcal{C}$ of the entire world concepts. Then, we introduce a stochastic model to describe the relevance of concepts to an action:
\begin{align}
\label{eq:conceptualizer}
p_{\scriptscriptstyle \mathbf{X}|A}(\vb*{x}|a;t)= \prod_{c\in\mathcal{C}}p_{\scriptscriptstyle X_c|A}(x_c|a;t), \end{align}
for all $\vb*{x}=(x_1,x_2,\dots,x_{|\mathcal{C}|}) \in \{\mathtt{TRUE},\mathtt{FALSE}\}^{|\mathcal{C}|}$.
The left-hand-side (LHS) in \eqref{eq:conceptualizer} states that the model is a conditional probability distribution of $\mathbf{X}$ given $a \in \mathcal{A}$, where $\mathbf{X} = (X_1,\dots,X_{|\mathcal{C}|})$ is the $|\mathcal{C}|$-tuple random variables of concept relevance indicators, and $A \in \mathcal{A}$ is the random variable of an intended action. The model is parameterized by an agent's task $t\in\mathcal{T}$, reflecting the variation of conceptualization across agents. We consider that the relevance of concepts to a given intended action is conditionally independent, yielding the right-hand-side (RHS) of \eqref{eq:conceptualizer}.\footnote{In real ML applications, the stochastic model about the action-concept relevance can be seen as a stochastic soft-decision such as the normalized logits of the cross entropy loss function in classification or the Kullback–Leibler divergence in knowledge distillation \cite{Hinton2015,Seo2020}, as well as the likelihood of a decision such as the stochastic policy in reinforcement learning (RL) \cite{Sutton2018}.}

\subsubsection{Action-Concept Model}
Consider a stochastic model of choosing a single symbolized concept $C$ that can convey an intention $A  = a$, where $C  \in \mathcal{C}$ is a random variable that depicts the chosen concept. Such a model, termed an \emph{action-to-concept (A2C) model}, can be described as
\begin{align}
\label{eq:A2C_new}p_{\scriptscriptstyle C|A}(c|a;t)= \frac{p_{\scriptscriptstyle X_c|A}(\mathtt{TRUE}|a;t)}{\sum_{c\in\mathcal{C}}p_{\scriptscriptstyle X_c|A}(\mathtt{TRUE}|a;t)}.
\end{align}
Note that \eqref{eq:A2C_new} is a normalized probability distribution of a concept $C$ being relevant to a given action $a$.

Moreover, given the model \eqref{eq:A2C_new}, a \emph{concept-to-action (C2A) model} can be readily obtained by using the Bayes rule:
\begin{align}
\label{eq:C2E}p_{\scriptscriptstyle A|C}(a|c;t)=\frac{p_{\scriptscriptstyle C|A}(c|a;t) p_{\scriptscriptstyle A}(a)}{\sum_{a\in\mathcal{A}}p_{\scriptscriptstyle C|A}(c|a;t) p_{\scriptscriptstyle A}(a)},
\end{align}
for all $a \in \mathcal{A}$, where $p_{\scriptscriptstyle A}(a)$ is the prior distribution of an intended action $A$ at the speaker.

\subsubsection{Concept-Symbol Model}
Even if the agents know the same concept, its representation at each agent may differ when they are developed in isolation, especially when the agents have learned the concept without supervision. A similar situation happens in humans, for example, when two people thinking of a concept `rabbit' might imagine different breed of rabbits even though they are conceptually the same. In this respect, a \emph{concept-to-symbol (C2S) model} is instrumental in harmonizing the agents by synchronizing their representation of concepts for semantic communication. To develop the C2S, there are two main issues that need to be addressed. First, a set of symbols $\mathcal{S}$ that are commonly known by the agents must be predetermined or emerged among the agents. Second, a C2S $s:\mathcal{C} \rightarrow \mathcal{S}$ that maps concepts to symbols should be predetermined or emerged among the agents. Alternatively, a centralized unit can harmonize the agents by carefully designing the symbol set and communication protocol. However, if the agents must learn them in a distributed manner, the emergence of both symbol set and communication protocol falls under the framework of emergent communication \cite{Lazaridou2020}.

{Throughout this work, we presume that there exists C2S, which is a deterministic one-to-one mapping $s:\mathcal{C}\rightarrow\mathcal{S}$ from the concept to the symbol set $\mathcal{S}$, where $s(c) \in \mathcal{S}$ describes the symbol that represents the concept $c$, while if $c \neq c'$ then $s(c) \neq s(c')$ and thereby $|\mathcal{C}| = |\mathcal{S}|$. However, we clarify that the framework proposed in this study is applicable even if C2S is not a one-to-one mapping, but many-to-one or one-to-many.} On the other hand, since the mapping is defined to be one-to-one, we have a deterministic \textit{symbol-to-concept (S2C)} model $s^{-1}:\mathcal{S}\rightarrow\mathcal{C}$, which is the inverse function of the C2S such that $s^{-1}(s(c)) = c$.

Overall, given an intended action $a \in \mathcal{A}$, a set of symbols that are obtained using the A2C and C2S models is referred to as a \emph{semantic representation (SR)} of $a$. Regarding the obtained SR as a source, traditional Shannon coding is applied after System 1 semantic coding in System 1 SNC, as illustrated in Fig. \ref{Fig_Shannon_semantic}.

\subsection{Shannon Coding under System 1 Semantic Coding}\label{subsec:shannon_communication}
The obtained SR, considered as a source, is first encoded by a source encoder followed by a channel encoder for efficient and effective communication.\footnote{We posit that source and channel codings are separately designed without loss of asymptotic optimality by assuming both the source and channel are discrete and memoryless \cite{Shannon1948,Vembu1994}.}
Specifically, the length of the binary uniquely decodable source-coded SR of an intended action quantifies the size of the SR in bits. The expected bit-length of SR in System 1 SNC can be derived as follows.
\begin{theorem} (Bit-Length of SR in System 1 SNC)\label{prop:ShannonSNC}
The expected bit-length of SR in System 1 SNC between agents with the same state is lower bounded as
\begin{align}\label{eq:lowerboundSNC}
    \mathsf{L}_{\text{S$_1$}} \geq -\sum_{c \in \mathcal{\mathcal{C}}} p_{\scriptscriptstyle X_{c}}(\mathtt{TRUE}) \log_2 \frac{p_{\scriptscriptstyle X_c}(\mathtt{TRUE})}{\sum_{c\in\mathcal{C}}p_{\scriptscriptstyle X_{c}}(\mathtt{TRUE})},
\end{align}
and upper bounded as
\begin{align}\label{eq:upperboundSNC}
    \mathsf{L}_{\text{S$_1$}} \leq -\sum_{c \in \mathcal{\mathcal{C}}} p_{\scriptscriptstyle X_{c}}(\mathtt{TRUE}) \left\lceil{\log_2 \frac{p_{\scriptscriptstyle X_c}(\mathtt{TRUE})}{\sum_{c\in\mathcal{C}}p_{\scriptscriptstyle X_{c}}(\mathtt{TRUE})}}\right\rceil,
\end{align}
where $p_{\scriptscriptstyle X_c}(\mathtt{TRUE}) = \sum_{a\in\mathcal{A}}p_{\scriptscriptstyle X_c|A}(\mathtt{TRUE}|a)p_{\scriptscriptstyle A}(a)$ is the probability of the concept $c$ being relevant to an action $a \in \mathcal{A}$, and $p_{\scriptscriptstyle A}$ is the prior distribution of an intended action $A\in\mathcal{A}$.
\end{theorem}
\begin{IEEEproof}
The proof is provided in Appendix \ref{appendix:proofofProp1}.
\end{IEEEproof}
Note that the task-specific agent state parameter is omitted for brevity in the above theorem under the premise that the communicating agents are in the same state. The theorem states that the SR bit-length depends on the number of concepts extracted from the intended action as well as the relative frequency of each concept's extraction among all actions. Meanwhile, although lossless source coding achieves both sufficiency and minimality of the SR in System 1 SNC over a noiseless channel, it lacks sufficiency in terms of communication accuracy under channel noise and fading. To overcome this, a channel encoder generates channel codes that achieve sufficiency by compromising minimality to ensure reliable communication of the source-coded SR. At the listener, the received channel code is successively decoded with the channel decoder and source decoder, to obtain the original symbolized concepts, followed by semantic decoding.
\section{Self Semantics-Native Communication via Contextual Reasoning}\label{sec:RHSC} In human communication, speakers are rational and self-conscious as to what they sound like, and change the ways they talk depending on the context that themselves and the listeners are involved in. In linguistics, this process is described as \emph{contextual reasoning} about the semantics in the local context of social interactions \cite{Bell1995, Bell1999, Bell2001}. Inspired by this, in this section we infuse contextual reasoning into System~1~SNC, and develop System 2 SNC. In System 2 semantic coding, before communicating, each agent performs contextual reasoning that is equivalent to \emph{self-SNC} with a virtual agent that mimics its listener. The self-SNC procedure sifts through all data semantics yielding the most-effective SR for its (physical) listener, improving communication efficiency, as we shall describe in the following.

\begin{figure*}[!t]
	\centering
	\includegraphics[width=0.6\textwidth]{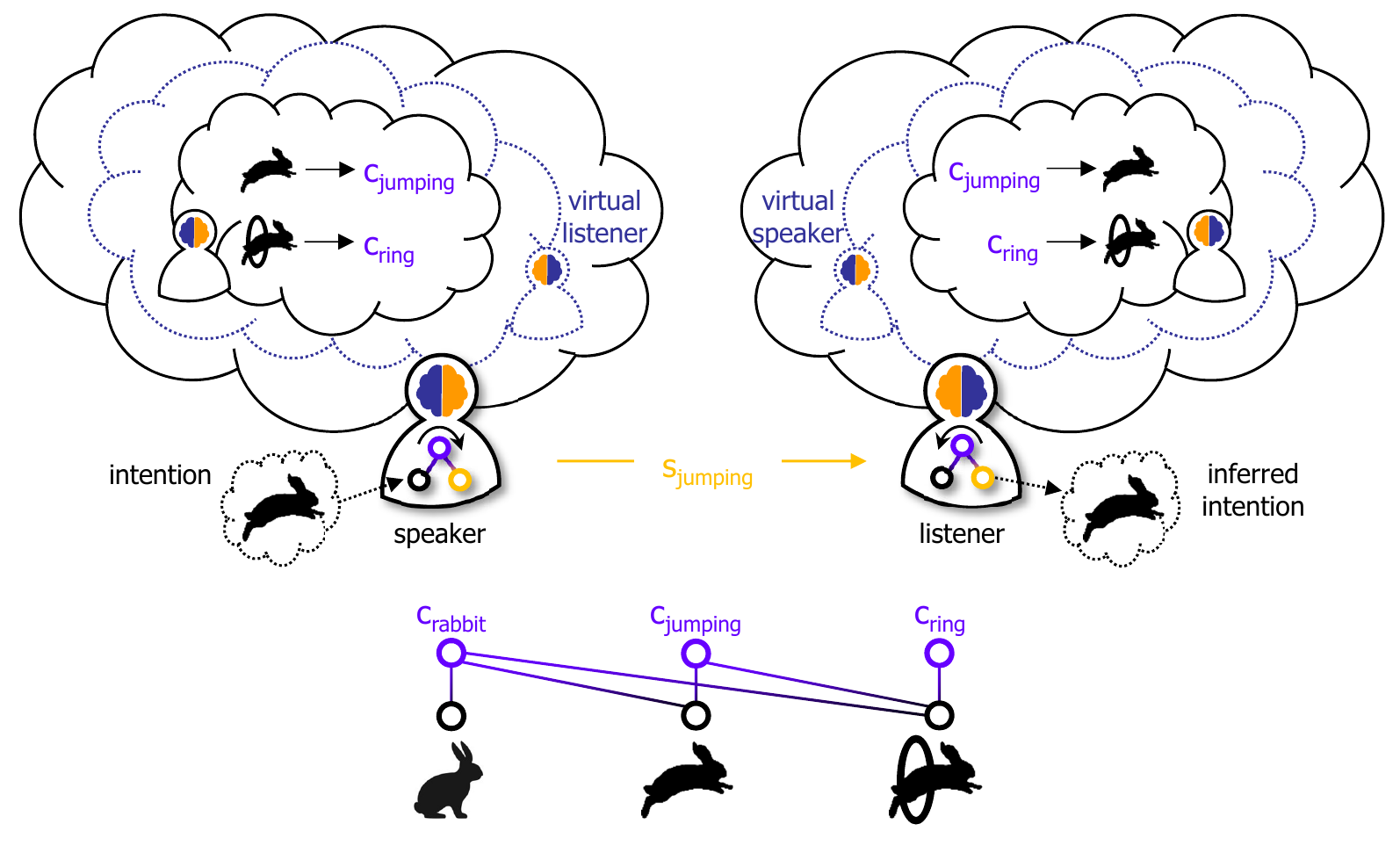}
	\caption{A rabbit referential game example with three types of actions having different concepts.}
	\label{fig:example}
	\vspace{-15pt}
\end{figure*}

\subsection{Contextual Reasoning}
In linguistics, contextual reasoning is often computationally described using the RSA model \cite{Frank2012, Goodman2013, Goodman2016, Frank2016}. The RSA model is rooted in the Gricean view of language use \cite{Grice1975}, presuming that people are `rational' agents who can communicate effectively and efficiently based on reasoning. In a similar vein, the rationality of System 2 SNC lies in the belief that agents use to reason about each other for more effective and efficient SNC. To illustrate the importance of contextual reasoning, we begin by providing a motivating example of System 2 SNC.

\subsubsection{A Referential Game Example}
Consider an instance of a world with a speaker and listener. Suppose there are three different actions that the listener can take: pointing out to an image of a rabbit sitting ($a_1$), rabbit jumping ($a_2$), and rabbit jumping into a ring ($a_3$). In addition, there exists a set $\mathcal{C} = \{c_\text{rabbit}$, $c_\text{jumping},c_\text{ring}\}$ of the three atomic concepts that are equivalently developed at each agent as illustrated in Fig.~\ref{fig:example}. Consider a communication-limited environment where the speaker must select only one of the given symbolized concepts, which are $\mathcal{S} = \{s_\text{rabbit}$, $s_\text{jumping}$, $s_\text{ring}$\}, to refer to an 
intended action to the listener. Suppose that the speaker induces the listener to take $a_2$, and intends to communicate it to the listener. Under the communication-limited environment, a na\"ive (System 1) speaker who should communicate all the symbolized concepts cannot describe $a_2$ to its listener.

By contrast, a rational (System 2) speaker, assuming its listener is also rational, would select the symbol $s_\text{jumping}$ to induce $a_2$. The rationale behind the selection is as follows. If the speaker communicates $s_{\text{rabbit}}$, then a rational listener will take $a_1$, since $c_{\text{rabbit}}$ is the only relevant concept of $a_1$, and thus $s_{\text{rabbit}}$ is the most efficient and effective representation of $a_1$. Likewise, if the speaker communicates $s_{\text{ring}}$, the listener will take $a_3$, since $c_{\text{ring}}$ is the unique concept of $a_3$. These two counter-examples justify the choice $s_\text{jumping}$ of the rational speaker. Meanwhile, upon receiving $s_{\text{jumping}}$, a na\"ive listener cannot identify which action the speaker refers to unless it additionally receives $s_{\text{rabbit}}$. A rational listener, by contrast, can directly take $a_2$ by reasoning in the same way as the rational speaker. In conclusion, the rational agents can exchange only a single symbol $s_{\text{jumping}}$ when referring to $a_2$, which significantly improves the communication efficiency without compromising accuracy.

\subsubsection{Contextual Reasoning via self-SNC}
As seen in the rabbit referential game example, seeking for the most meaningful concepts is crucial for improving the efficiency of SNC. One major challenge is that the meaningfulness of a concept is context-dependent, determined by its effectiveness in achieving a given goal of SNC. In the world of the rabbit referential game example, all entities are associated with the concept $c_{\text{rabbit}}$ that can be an effective concept for describing the entire world but are not meaningful for referring to a certain action. On the contrary, $c_{\text{ring}}$ is a unique concept that is effective in referring to $a_3$, but otherwise becomes meaningless. Another challenge comes from the fact that each agent understand different context, which is formed by complex factors such as the world configuration and the agent's knowledge and beliefs. In the SNC considered in this article, the individual context is affected by the agent's mapping between concepts and actions as well as by the prior beliefs about these concepts and actions, resulting in heterogeneous individual contexts for different agents.

Contextual reasoning can overcome the aforementioned difficulties as follows. Assuming that the states of all agents are known, each agent can obtain the individual context, based on which it can apply conditional reasoning. A rational speaker, as an example, can reason that `if I were the rational listener, I would have chosen $a_3$ upon listening to $s_{\text{ring}}$; therefore, I should communicate $s_{\text{ring}}$.' Similarly, a rational listener can reason that `if I were the rational speaker, I would have communicated $s_{\text{jumping}}$ to describe $a_2$; therefore, I will select $a_2$ when listening to $s_{\text{jumping}}$.' Such contextual reasoning can be seen as a self-SNC between a rational agent and a virtual communication counterpart built on the individual context of its rational listener. While iterating this self-SNC, the individual contexts of the agent and its virtual counterpart converge towards a focal point, referred to as a \emph{mutual context}. The converged mutual context gives rise to the emergent A2C and C2A models, i.e., System 2 semantic coding as will be elaborated in the following subsection.


\subsection{System 2 Semantic Coding}\label{subsec:RCRD}
System 2 semantic coding is enabled by a \emph{rational action-to-concept (rA2C) model} and a \emph{rational concept-to-action (rC2A) model} that emerge through self-SNC.
As opposed to A2C and C2A, rA2C and rC2A include the self-SNC which is carried out locally at each agent, such that their stochastic models are parameterized by the states of both speaker and listener, as detailed next.

Consider a speaker in a task-specific state $t \in \mathcal{T}$ and a listener in $t' \in \mathcal{T}$, and assume that both of them know each other's states. For a given intended action $a\in\mathcal{A}$, the rA2C is denoted by $
p_{\scriptscriptstyle C|A}(c|a;\mathbf{t})$ for all $c \in \mathcal{C}$, where $\mathbf{t} = (t,t')$ denotes the tuple of agents states. The above rA2C is a conditional probability distribution of choosing $c \in \mathcal{C}$ as a meaningful concept to represent an intended action $a \in \mathcal{A}$. On the other hand, for a given meaningful concept $c\in\mathcal{C}$, the rC2A is denoted by $p_{\scriptscriptstyle A|C}(a|c;\mathbf{t})$ for all $a\in\mathcal{A}$. The rC2A is a conditional probability distribution for inferring the intended action $A$ from the meaningful concept $C$.

In what follows, we show how to obtain the optimal rA2C and rC2A by self-SNC. Recall that the individual context is affected by the mapping between concepts and actions, as well as the prior beliefs. Thus, the individual context of a speaker can be described as 
\begin{align}\label{eq:speakercontext}
\mathsf{S}(a,c;\mathbf{t}) = \frac{p_{\scriptscriptstyle C|A}(c|a;\mathbf{t}) p_{\scriptscriptstyle A}(a)}{\sum_{a\in\mathcal{A}}\sum_{c\in\mathcal{C}}p_{\scriptscriptstyle C|A}(c|a;\mathbf{t}) p_{\scriptscriptstyle A}(a)},
\end{align}
for all $(a,c)\in\mathcal{A}\times\mathcal{C}$. Note that \eqref{eq:speakercontext} is a joint probability distribution of concepts and actions parameterized by agents' states, which is a normalized function of rA2C and prior distribution $p_{\scriptscriptstyle A}(a)$ of actions for all $a\in\mathcal{A}$. Similarly, the individual context of a listener is described as a function of rC2A and prior distribution $p_{\scriptscriptstyle C}(c)$ of concepts for all $c\in\mathcal{C}$, i.e., 
\begin{align}\label{eq:listenercontext}
\mathsf{L}(a,c;\mathbf{t}) = \frac{p_{\scriptscriptstyle A|C}(a|c;\mathbf{t}) p_{\scriptscriptstyle C}(c)}{\sum_{a\in\mathcal{A}}\sum_{c\in\mathcal{C}}p_{\scriptscriptstyle A|C}(a|c;\mathbf{t}) p_{\scriptscriptstyle C}(c)},
\end{align}
for all $(a,c)\in\mathcal{A}\times\mathcal{C}$. Both individual contexts are normalized to be in a form of joint distribution of an action $A\in\mathcal{A}$ and meaningful concept $C\in\mathcal{C}$ parameterized by $t$ and $t'$.

Now, denote by $\mathsf{M}(a,c;\mathbf{t})$ for all $(a,c)\in\mathcal{A}\times\mathcal{C}$, the mutual context of the speaker and listener and suppose that the speaker and listener independently and individually minimize an objective function defined by
\begin{align}\label{eq:lossfunction}
\mathsf{G} = \lambda\! \left[ \mathsf{H}(\mathsf{S},\mathsf{M}) \! - \! \frac{\mathsf{H}(\mathsf{S})}{\alpha} \right]\! +\! (1\!-\!\lambda) \!\left[ \mathsf{H}(\mathsf{L},\mathsf{M}) \!-\! \frac{\mathsf{H}(\mathsf{L})}{\beta} \right],
\end{align}
with respect to both individual contexts $\mathsf{S}$ and $\mathsf{L}$, and mutual context $\mathsf{M}$, given parameters $0 < \lambda < 1$ and $\alpha,\beta \geq 1$. In \eqref{eq:lossfunction}, $\mathsf{H}(\mathsf{S}) = -\sum_{(a,c)}\mathsf{S}(a,c;\mathbf{t})\log \mathsf{S}(a,c;\mathbf{t})$ and $\mathsf{H}(\mathsf{L}) = -\sum_{(a,c)}\mathsf{L}(a,c;\mathbf{t})\log \mathsf{L}(a,c;\mathbf{t})$ are the joint entropies of the action and meaningful concept given the agent's state with respect to the individual contexts $\mathsf{S}$ and $\mathsf{L}$, respectively. The term $\mathsf{H}(\mathsf{S},\mathsf{M})  -\sum_{(a,c)}\mathsf{S}(a,c;\mathbf{t})\log \mathsf{M}(a,c;\mathbf{t})$ is the cross entropy of $\mathsf{M}$ and $\mathsf{S}$, and $\mathsf{H}(\mathsf{L},\mathsf{M}) = -\sum_{(a,c)}\mathsf{L}(a,c;\mathbf{t})\log \mathsf{M}(a,c;\mathbf{t})$ is the cross entropy of $\mathsf{M}$ and $\mathsf{L}$.

To illustrate the intuitive meaning of the minimization of \eqref{eq:lossfunction}, first consider $\alpha = \beta = 1$, which reduces \eqref{eq:lossfunction} to a weighted sum of two KL-divergences, one between $\mathsf{S}$ and $\mathsf{M}$ and the other between $\mathsf{L}$ and $\mathsf{M}$, such that
\begin{align}\label{eq:lossfunctionKL}
\mathsf{G}_{\scriptscriptstyle \alpha,\beta = 1} = \lambda\mathsf{D}_\text{KL}(\mathsf{S}||\mathsf{M}) + (1-\lambda)\mathsf{D}_\text{KL}(\mathsf{L}||\mathsf{M}).
\end{align}
Minimizing \eqref{eq:lossfunctionKL} in terms of $\mathsf{S}$ and $\mathsf{L}$, given $\mathsf{M}$, reduces the divergence between two distributions $\mathsf{S}$ and $\mathsf{L}$, and make them move towards a focal point $\mathsf{M}$. Moreover, for fixed $\mathsf{S}$ and $\mathsf{L}$, minimizing \eqref{eq:lossfunctionKL} in terms of $\mathsf{M}$ finds the division point on the line segment between $\mathsf{S}$ and $\mathsf{L}$ (see Appendix \ref{appendix:proofofAM}). Meanwhile, minimizing \eqref{eq:lossfunctionKL} induces the maximization of both $\mathsf{H}(\mathsf{S})$ and $\mathsf{H}(\mathsf{L})$. For the fixed prior distributions $p_{\scriptscriptstyle A}(a)$ and $p_{\scriptscriptstyle C}(c)$, for all $a\in \mathcal{A}$ and $c\in \mathcal{C}$, respectively, maximizing $\mathsf{H}(\mathsf{S})$ results in increasing the uncertainty of conceptualizing the rA2C and maximizing $\mathsf{H}(\mathsf{L})$ results in increasing the rC2A uncertainty. Thus, both of them are closely related to the performance of System 2 SNC, in that the maximization of the former reduces communication efficiency by increasing the lower bound of the expected bit-length of SR (see Section \ref{subsec:shannon_communication}), while the maximization of the latter reduces the deconceptualization accuracy.

Therefore, such factors are controlled by hyperparameters $\alpha$ and $\beta$ in \eqref{eq:lossfunction}, which determine how rational the rA2C and rC2A are. For example, setting $\alpha > 1$ promotes representational efficiency by suppressing the maximization of $\mathsf{H}(\mathsf{S})$, while setting $\beta >1$ promotes deconceptualization accuracy by suppressing the maximization of $\mathsf{H}(\mathsf{L})$. However, setting $\alpha$ and $\beta$ to be too large does not always make  System 2 SNC efficient and effective, since large $\alpha$ may result in obtaining rA2C that maps multiple different actions to the same meaningful concept, and large $\beta$ may result in obtaining rC2A that infers the same action from multiple different meaningful concepts. Numerical experiments illustrating this aspect are shown in Section \ref{sec:experiments}.

The minimization of \eqref{eq:lossfunction} is a variational problem, which can be solved by alternating minimization with respect to $\mathsf{S}$, $\mathsf{L}$ and $\mathsf{M}$, in the order of $\mathsf{M} \!\rightarrow\! \mathsf{S} \!\rightarrow\! \mathsf{M} \!\rightarrow\! \mathsf{L} \!\rightarrow\! \mathsf{M} \!\rightarrow\! \cdots$ as formalized in Theorem \ref{thm:theonly}.
\begin{theorem} (Mutual Context Convergence) \label{thm:theonly} 
As the recursion depth $r\rightarrow \infty$, the alternating iterations~of
\begin{align}
\label{eq:iteration_1}\mathsf{M}^{\scriptscriptstyle [r]}_{\scriptscriptstyle 1}(a,c;\mathbf{t}) &= \lambda \mathsf{S}^{\scriptscriptstyle [r-1]}(a,c;\mathbf{t}) \! + \! (1\!-\!\lambda)\mathsf{L}^{\scriptscriptstyle [r-1]}(a,c;\mathbf{t}),\\
\label{eq:iteration_2}\mathsf{S}^{\scriptscriptstyle [r]}(a,c;\mathbf{t}) &= \frac{\mathsf{M}^{\scriptscriptstyle [r]}_{\scriptscriptstyle 1}(a,c;\mathbf{t})^\alpha}{\sum_{(a,c)\in\mathcal{A}\times\mathcal{C}}\mathsf{M}^{\scriptscriptstyle [r]}_{\scriptscriptstyle 1}(a,c;\mathbf{t})^\alpha},
\\
\label{eq:iteration_3}\mathsf{M}^{\scriptscriptstyle [r]}_{\scriptscriptstyle 2}(a,c;\mathbf{t}) &= \lambda \mathsf{S}^{\scriptscriptstyle [r]}(a,c;\mathbf{t}) \! + \! (1\!-\!\lambda)\mathsf{L}^{\scriptscriptstyle [r-1]}(a,c;\mathbf{t}),\\
\label{eq:iteration_4}\mathsf{L}^{\scriptscriptstyle [r]}(a,c;\mathbf{t}) &= \frac{\mathsf{M}^{\scriptscriptstyle [r]}_{\scriptscriptstyle 2}(a,c;\mathbf{t})^\beta}{\sum_{(a,c)\in\mathcal{A}\times\mathcal{C}}\mathsf{M}^{\scriptscriptstyle [r]}_{\scriptscriptstyle 2}(a,c;\mathbf{t})^\beta}
\end{align}
converge to a mutual context $\mathsf{M}^{\scriptscriptstyle [*]} = \lim_{r\rightarrow\infty}\mathsf{M}_{\scriptscriptstyle 1}^{\scriptscriptstyle [r]} = \lim_{r\rightarrow\infty}\mathsf{M}_{\scriptscriptstyle 2}^{\scriptscriptstyle [r]}$ for all $(a, c) \in \mathcal{A}\times\mathcal{C}$, which is a local minimum of \eqref{eq:lossfunction}.
\end{theorem}
\begin{IEEEproof}
The proof is provided in Appendix \ref{appendix:proofofAM}.
\end{IEEEproof}
In other words, Theorem \ref{thm:theonly} states that
\begin{align}
\mathsf{G}(\mathsf{S}^{\scriptscriptstyle [r-1]}, \mathsf{L}^{\scriptscriptstyle [r-1]}) \geq \mathsf{G}(\mathsf{S}^{\scriptscriptstyle [r]}, \mathsf{L}^{\scriptscriptstyle [r-1]}) \geq \mathsf{G}(\mathsf{S}^{\scriptscriptstyle [r]}, \mathsf{L}^{\scriptscriptstyle [r]}),
\end{align}
for all recursion depth $r\geq 1$. Moreover, the solution of Theorem 2 is locally optimal, since \eqref{eq:lossfunction} is not jointly convex with respect to $\mathsf{S}$, $\mathsf{L}$ and $\mathsf{M}$. For the special case when $\alpha=\beta=1$, we show the global optimality of the solutions, as elaborated next.

{
The boundary condition of the recursive rA2C and rC2A at depth $r= 0$ are given by A2C and C2A of System 1 SNC, respectively, i.e.,
\begin{align}\label{eq:initialrE2C}
p^{\scriptscriptstyle [0]}_{\scriptscriptstyle C|A}(c|a;\mathbf{t}) &= \frac{p_{\scriptscriptstyle X_c|A}(\mathtt{TRUE}|a;t)}{\sum_{c\in\mathcal{C}}p_{\scriptscriptstyle X_c|E}(\mathtt{TRUE}|a;t)}\ \ \text{and}\\
\label{eq:initialrC2E}
p^{\scriptscriptstyle [0]}_{\scriptscriptstyle A|C}(a|c;\mathbf{t}) &= \frac{p_{\scriptscriptstyle A|X_c}(a|\mathtt{TRUE};t')}{\sum_{a\in\mathcal{A}}p_{\scriptscriptstyle A|X_c}(a|\mathtt{TRUE};t')},
\end{align}
for all $c \in \mathcal{C}$ and $a\in\mathcal{A}$, respectively, similar to \eqref{eq:A2C_new} and \eqref{eq:C2E}. Substituting \eqref{eq:initialrE2C} and \eqref{eq:initialrC2E} respectively into the individual contexts \eqref{eq:speakercontext} and \eqref{eq:listenercontext} gives $\mathsf{S}^{\scriptscriptstyle [0]}$ and $\mathsf{L}^{\scriptscriptstyle [0]}$ when initializing \eqref{eq:iteration_1} to \eqref{eq:iteration_4}.} Denote by $\mathsf{G}^{\scriptscriptstyle [*]} = \mathsf{G}(\mathsf{S}^{\scriptscriptstyle [*]},\mathsf{L}^{\scriptscriptstyle [*]})$ the stationary point of \eqref{eq:lossfunction} at the found minimum, and $\mathsf{S}^{\scriptscriptstyle [*]} = \lim_{r\to\infty}\mathsf{S}^{\scriptscriptstyle [r]}$, and $\mathsf{L}^{\scriptscriptstyle [*]} = \lim_{r\to\infty}\mathsf{L}^{\scriptscriptstyle [r]}$ the stationary points of $\mathsf{S}$ and $\mathsf{L}$, respectively. Then, the mutual context convergence in Theorem 2 can be recast as the individual context convergence.
\begin{corollary}\label{cor:proofofPequalsQ}
(Individual Context Convergence) For any parameters $\alpha, \beta \geq 1$ and $0 < \lambda < 1$,
\begin{align}
    \mathsf{S}^{\scriptscriptstyle [*]}(a,c;\mathbf{t}) = \mathsf{L}^{\scriptscriptstyle [*]}(a,c;\mathbf{t}) = \mathsf{M}^{\scriptscriptstyle [*]}(a,c;\mathbf{t})
\end{align}
holds for all $(a,c)\in\mathcal{A}\times\mathcal{C}$.
\end{corollary}
\begin{IEEEproof}
The proof is provided in Appendix \ref{appendix:proofofPequalsQ}.
\end{IEEEproof}

\begin{remark}(Global Optimality) For $\alpha=\beta=1$, the loss function \eqref{eq:lossfunction} of Theorem 2 boils down to \eqref{eq:lossfunctionKL} that can be minimized when all the KL-divergence terms become zero since the KL-divergence is non-negative. The solution of Theorem 2 achieves this result according to Corollary \ref{cor:proofofPequalsQ}, which is thus the global minimum.
\end{remark}
\noindent With the solution of Theorem 2, note also that \eqref{eq:lossfunctionKL} is the $\lambda$-divergence between $\mathsf{S}^{\scriptscriptstyle [*]}$ and $\mathsf{L}^{\scriptscriptstyle [*]}$. For $\lambda=0.5$, \eqref{eq:lossfunctionKL} yields the Jensen-Shannon divergence whose minimum is zero, which can be achieved by the solution of Theorem 2.

\begin{figure*}
\centering
\subfigure[]{\includegraphics[width=0.45\textwidth]{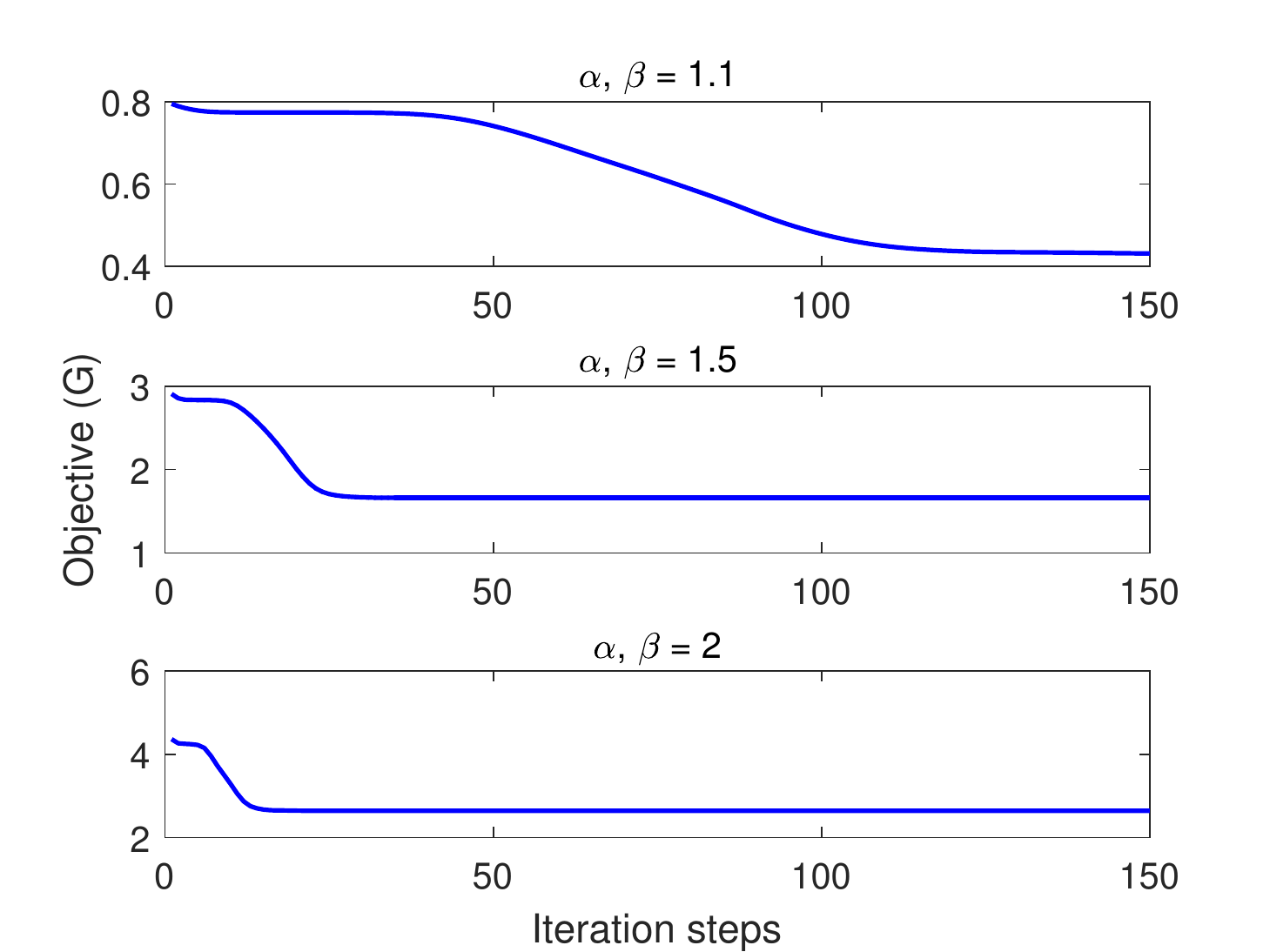}\label{fig:iteration}}
\subfigure[]{\includegraphics[width=0.45\textwidth]{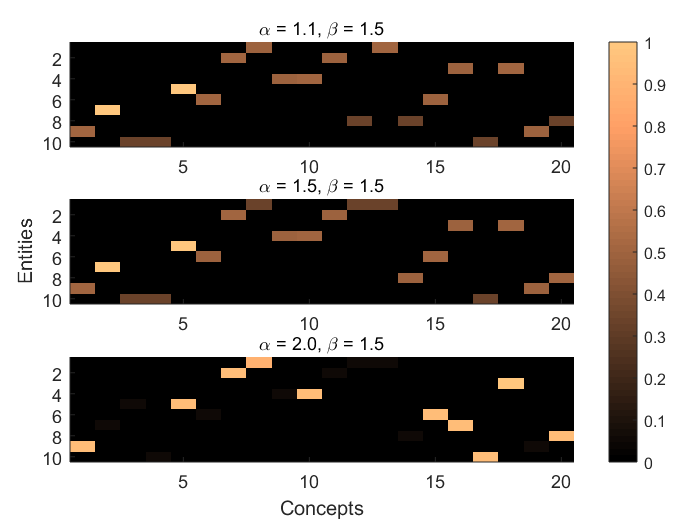}\label{fig:heatmap}}
\caption{Illustrations of (a) convergence of $\mathsf{G}$ through the recursion of rA2C and rC2A with different parameters $\alpha, \beta = 1.1, 1.5\text{ and } 2$, for some instance of the world with fixed $|\mathcal{E}| = |\mathcal{C}| = 100$ and $\lambda = 0.5$; (b) empirical distribution (raw-stochastic) of chosen meaningful concept for a given entity based on the stationary rA2C, with different parameters $\alpha = 1.1, 1.5 \text{ and } 2.0$, and fixed $|\mathcal{E}| = 10$, $|\mathcal{C}|=20$, $\beta = 1.5$, $\lambda = 0.5$.}
\end{figure*}



Next, the alternating iterations of \eqref{eq:iteration_1}-\eqref{eq:iteration_4} in Theorem 2 allow us to unravel the self-SNC operations between an agent and its virtual interlocutor as follows. 
\begin{corollary}
({Self-SNC}) By recasting \eqref{eq:iteration_1}-\eqref{eq:iteration_4}, we can derive the rA2C at iteration step $t\geq 1$
\begin{align}\label{eq:RC}
p_{\scriptscriptstyle C|A}^{\scriptscriptstyle [r]}(c|a;\mathbf{t}) &= \frac{\mathsf{M}^{\scriptscriptstyle[r]}_{\scriptscriptstyle 1}(a,c;\mathbf{t})^\alpha }{\sum_{\forall c \in \mathcal{C}} \mathsf{M}^{\scriptscriptstyle[r]}_{\scriptscriptstyle 1}(a,c;\mathbf{t})^{\alpha}}
\end{align}
and the rC2A at the recursion depth $r$
\begin{align}\label{eq:RD}
p_{\scriptscriptstyle A|C}^{\scriptscriptstyle [r]}(a|c;\mathbf{t}) &= \frac{\mathsf{M}^{\scriptscriptstyle [r]}_{\scriptscriptstyle 2}(a,c;\mathbf{t})^\beta}{\sum_{\forall a \in \mathcal{A}} \mathsf{M}^{\scriptscriptstyle [r]}_{\scriptscriptstyle 2}(a,c;\mathbf{t})^\beta},
\end{align}
for all $a \in \mathcal{A}$, $c \in \mathcal{C}$ given $\mathbf{t} = (t,t') \in \mathcal{T}^2$, which minimizes \eqref{eq:lossfunction} when $r\rightarrow \infty$.
\end{corollary}
The obtained rA2C can be seen as a scoring function of the context-dependent meaningfulness of concepts for communicating an entity. On the other hand, the rC2A scores the correctness of the inferred intended entity given a meaningful concept. Furthermore, \eqref{eq:RC} and \eqref{eq:RD} form an iterative recursion, which implements the self-SNC between an agent and its virtual interlocutor as described earlier. An infinite recursion $r\rightarrow \infty$ of \eqref{eq:RC} and \eqref{eq:RD}, induces their convergence, as well as the objective function \eqref{eq:lossfunction} to a stationary point $\mathsf{G}^{\scriptscriptstyle [*]}$. For simplicity, the notations $p_{\scriptscriptstyle C|A}(c|a;\mathbf{t})$ for all $c\in\mathcal{C}$ and $p_{\scriptscriptstyle A|C}(a|c;\mathbf{t})$ for all $a\in\mathcal{A}$ are now regarded as the stationary rA2C and rC2A, respectively, at the convergence of $\mathsf{G}^{\scriptscriptstyle [*]}$, unless stated otherwise. Experimental results showing the convergence dynamics of $\mathsf{G}$ with respect to the iteration step of the recursion are illustrated in Fig. \ref{fig:iteration} with different settings of the rationality parameters $\alpha, \beta > 1$. We find that the objective \eqref{eq:lossfunction} converges after some iterations, and the speed of convergence is faster for larger $\alpha$ and $\beta$. However, as mentioned earlier, large values of $\alpha$ and $\beta$, does not always provide good solutions for reliable System 2 SNC; thus $\alpha$ and $\beta$ should be carefully chosen.

\begin{remark}
As a special case, the  rA2C and rC2A recursion can be recast as an iterative matrix scaling method termed Sinkhorn scaling \cite{Sinkhorn1964,Knopp1967} with parameters $\alpha = 1$, $\beta =1$ and $\lambda = 1$. Since the Sinkhorn scaling suboptimizes the optimal transport (or earth moving) problem \cite{Cuturi2013}, the recursion can be viewed as solving an optimal transport problem between the two priors $p_{\scriptscriptstyle A}$ and $p_{\scriptscriptstyle C}$, in which  the rA2C is cast as a transport plan from an action to a concept and rC2A as that from a concept to an action \cite{Wang2020}, which is basically an infinite recursion of the RSA model \cite{Frank2012, Goodman2013, Goodman2016, Frank2016}.
\end{remark}

\subsection{Reliable System 2 SNC with Multiple Meaningful Concepts}\label{sec:sequentialRHSC}
The above-mentioned self-SNC based approach selects a single meaningful concept at a time, however, communicating multiple meaningful concepts can improve the reliability of System 2 SNC. To this end, there are two main directions: a planning-based method that selects multiple meaningful concepts at once, and a greedy algorithm-based method that selects meaningful concepts one-by-one. The former may give better performance based on optimal planning, however, the algorithm is computationally expensive since the selection of multiple meaningful concepts should be jointly designed. Moreover, it is hard to find the minimum number of concepts  guaranteeing reliable communication, thereby calling for a full search to obtain the optimal planning. On the other hand, the latter is computationally cheap, since one meaningful concept is selected at a time and the algorithm can stop whenever the communication reliability is guaranteed.

Therefore, we consider the latter based on a greedy algorithm, which boils down to a problem of updating a pair of rA2C and rC2A after every meaningful concept selection and communication. In brief, since the meaningful concepts communicated in the past affect both the beliefs (priors) of the speaker and listener about the intended entity and meaningful concepts in the present, rA2C and rC2A should be updated based on self-SNC under the updated beliefs. To illustrate, consider a speaker in a state $t\in\mathcal{T}$ selecting two meaningful concepts $c_1, c_2 \in \mathcal{C}$ in sequence to communicate with a listener in state $t' \in \mathcal{T}$. Suppose the stationary rA2C is obtained via self-SNC with initial priors $p_{\scriptscriptstyle A}(a)$, $\forall a\in\mathcal{A}$ and $p_{\scriptscriptstyle C}(c)$, $\forall c\in\mathcal{C}$. The first meaningful concept is chosen by $c_1 = \argmax_c p_{\scriptscriptstyle C|A}(c|a;\mathbf{t})$, where $\mathbf{t} = (t,t')$, and communicated in the form of a symbolized concept $s(c_1)$. At the listener, upon receiving $s(c_1)$, the prior distribution about the intended action is updated by $p_{\scriptscriptstyle A}(a) \leftarrow p_{\scriptscriptstyle A|C}(a|c_1;\mathbf{t})$, $\forall a\in\mathcal{A}$. Furthermore, for the next meaningful concept selection, since $c_1$ henceforth is no longer meaningful, the prior distribution is updated by $p_{\scriptscriptstyle C}(c) \leftarrow \frac{p_{\scriptscriptstyle C}(c)}{\sum_{c\in \mathcal{C}\backslash c_1} p_{\scriptscriptstyle C}(c)}$, $\forall c \in \mathcal{C}\backslash c_1$ and $p_{\scriptscriptstyle C}(c_1) \leftarrow 0$. Then, rA2C and rC2A are also updated based on the updated prior distributions to select the next meaningful concept.

\begin{algorithm}[t]
\DontPrintSemicolon
  \algsetup{linenosize=\tiny}
  \small
  \KwInput{$\mathcal{A}$; $\mathcal{C}$; $p_{\scriptscriptstyle \mathbf{X}|A}(\vb*{x}|a;t)$, $p_{\scriptscriptstyle A|\mathbf{X}}(a|\vb*{x};t')$, $\forall{\vb*{x}}\in\{\mathtt{TRUE},\mathtt{FALSE}\}^{|\mathcal{C}|}$, $\forall a\in\mathcal{A}$; $p_{\scriptscriptstyle A}(a)$, $\forall a\in\mathcal{A}$; $p_{\scriptscriptstyle C}(c)$, $\forall c \in\mathcal{C}$, $\alpha$, $\beta$, $\lambda$}
  \KwOutput{$K$ meaningful concepts $c_1$,$c_2$,\dots,$c_K$}
  Fix an intended action $\hat{a}\in\mathcal{A}$ of the speaker\;
  \For{$k=1$ to $K$}
  {
  \KwInitialize{$\mathsf{S}(a,c;\mathbf{t}) \propto p_{\scriptscriptstyle X_c|E}(\mathtt{TRUE}|a;t)p_{\scriptscriptstyle A}(a)$, $\mathsf{L}(a,c;\mathbf{t}) = p_{\scriptscriptstyle A|X_c}(a|\mathtt{TRUE};t')p_{\scriptscriptstyle C}(c)$, $\forall (a,c)\in\mathcal{A}\times\mathcal{C}$;}
  \Repeat{convergence}
  {
  $\mathsf{M}(a,c;\mathbf{t}) \leftarrow \lambda \mathsf{S}(a,c;\mathbf{t}) + (1-\lambda) \mathsf{L}(a,c;\mathbf{t})$, $\forall (a,c)\in\mathcal{A}\times\mathcal{C}$\;
  $\mathsf{S}(a,c;\mathbf{t}) \leftarrow \frac{\mathsf{M}(a,c;\mathbf{t})^\alpha}{\sum_{(a,c)\in\mathcal{A}\times\mathcal{C}}\mathsf{M}(a,c;\mathbf{t})^\alpha}$, $\forall (a,c)\in\mathcal{A}\times\mathcal{C}$\;
  $\mathsf{M}(a,c;\mathbf{t}) \leftarrow \lambda \mathsf{S}(a,c;\mathbf{t}) + (1-\lambda) \mathsf{L}(a,c;\mathbf{t})$, $\forall (a,c)\in\mathcal{A}\times\mathcal{C}$\;
  $\mathsf{L}(a,c;\mathbf{t}) \leftarrow \frac{\mathsf{M}(a,c;\mathbf{t})^\beta}{\sum_{(a,c)\in\mathcal{A}\times\mathcal{C}}\mathsf{M}(a,c;\mathbf{t})^\beta}$, $\forall (a,c)\in\mathcal{A}\times\mathcal{C}$
  }
$p_{\scriptscriptstyle C|A}(c|a;\mathbf{t}) \leftarrow \frac{\mathsf{M}(a,c;\mathbf{t})^\alpha}{\sum_{c\in\mathcal{C}}\mathsf{M}(a,c;\mathbf{t})^\alpha}$\;
$p_{\scriptscriptstyle A|C}(a|c;\mathbf{t}) \leftarrow \frac{\mathsf{M}(a,c;\mathbf{t})^\alpha}{\sum_{a\in\mathcal{A}}\mathsf{M}(a,c;\mathbf{t})^\alpha}$\;
$c_k = \argmax_{c} p_{\scriptscriptstyle C|A}(c|\hat{a};\mathbf{t})$\;
$\mathcal{C}_k = \mathcal{C}_{k-1}\backslash c_{k}$ ($\mathcal{C}_0 = \mathcal{C})$\;
$p_{\scriptscriptstyle A}(a) \leftarrow p_{\scriptscriptstyle A|C}(a|c_k;\mathbf{t})$, $\forall a\in\mathcal{A}$\;
$p_{\scriptscriptstyle C}(c) \leftarrow \frac{p_{\scriptscriptstyle C}(c)}{\sum_{c\in \mathcal{C}_k} p_{\scriptscriptstyle C}(c)}$, $\forall c\in\mathcal{C}_k$, $p_{\scriptscriptstyle C}(c_k) \leftarrow 0$\;
}
\caption{Selecting $K$ Meaningful Concepts for System 2 SNC}
\label{algo:1}
\end{algorithm}

Likewise, $k \geq 1$ meaningful concepts can be chosen by sequentially obtaining rA2Cs and corresponding rC2As. The pseudo-code for obtaining sequential pairs of rA2C and rC2A is provided in Algorithm \ref{algo:1}. {Note that the (asymptotic) computational time complexity of Algorithm \ref{algo:1} is $\mathcal{O}(KR|\mathcal{A}|^2|\mathcal{C}|^2)$, where $K$ is the number of selecting meaningful concepts for System 2 SNC, $R$ is the number of iterations until convergence, and $\mathcal{A}$ and $\mathcal{C}$ are the sets of actions and concepts, respectively.} Meanwhile, Theorem \ref{thm:reliability} states that the accuracy of inferring the intended action by rC2A enhances as the number of communicated meaningful concepts increases.
\begin{theorem}\label{thm:reliability}
(Reliability Enhancement) In System 2 SNC, for a given intended action $a\in\mathcal{A}$, the probability of successfully inferring $a$ with the stationary rC2A is non-decreasing with the number of communication rounds, i.e.,
\begin{align}
    p^{k-1}_{\scriptscriptstyle A|C}(a|c_{k-1};\mathbf{t}) \leq p^{k}_{\scriptscriptstyle A|C}(a|c_{k};\mathbf{t})
\end{align}
for $k\geq 2$ and fixed parameters $\alpha,\beta\geq 1$ (excluding $\alpha=\beta = 1$) and $0<\lambda<1$ over communication rounds, where $p^{k}_{\scriptscriptstyle A|C}(a|c;\mathbf{t})$ and $c_k$ denote the $k$-th updated stationary rC2A and $k$-th meaningful concept selected with the $k$-th updated stationary rA2C, respectively.
\end{theorem}
\begin{IEEEproof}
The proof is provided in Appendix \ref{appendix:proofofreliability}.
\end{IEEEproof}
The above theorem proves that the greedy approach is correct in the sense that the greedy selection of meaningful concepts will eventually guarantee reliable communication.

\subsection{Shannon Coding under System 2 Semantic Coding}\label{subsec:shannon_communication_RHSC}
Let $K$ be the minimum number of meaningful concepts, i.e., $c_1,c_2,\dots,c_K$, that guarantees reliable communication in System 2 SNC. Then, the SR of an intended action is the collection of $K$ symbolized meaningful concepts. As described in Section \ref{subsec:shannon_communication}, the SR is encoded via source and channel coding to ensure minimality and sufficiency for the physical transmission over a noiseless/noisy channel. Here, the (lossless) source coding depends on the distribution of the meaningful concept, which is examined by the obtained stationary rA2C. To further illustrate, Fig. \ref{fig:heatmap} shows the empirical distribution of the meaningful concepts for a given intended action with a different rationality parameter $\alpha$. Note that with increasing $\alpha$, the uncertainty of meaningful concept for each action is reduced, i.e., $\mathsf{H}(\mathsf{S})$ in \eqref{eq:lossfunction} is reduced. Reflecting this, the expected bit-length of SR in System 2 SNC can be derived as follows.
\begin{corollary}\label{cor:3}
(Bit-Length of SR in System 2 SNC) The minimum expected bit-length of SR composed of $K$ symbolized meaningful concepts in System 2 SNC between a speaker in state $t\in\mathcal{T}$ and listener in state $t'\in\mathcal{T}$ is lower bounded as
\begin{align}\label{eq:irSNC_lower}
\mathsf{L}_{\text{S$_2$}}(\mathbf{t}) &\geq  -\sum_{k = 1}^{K}\sum_{c\in\mathcal{C}} p^k_{\scriptscriptstyle C}(c;\mathbf{t}) \log_2 p^k_{\scriptscriptstyle C}(c;\mathbf{t}),
\end{align}
where $\mathbf{t} = (t,t')$, and upper bounded as
\begin{align}\label{eq:irSNC_upper}
\mathsf{L}_{\text{S$_2$}}(\mathbf{t}) &\leq  \sum_{k = 1}^{K}\sum_{c\in\mathcal{C}} p^k_{\scriptscriptstyle C}(c;\mathbf{t}) \left\lceil -\log_2 p^k_{\scriptscriptstyle C}(c;\mathbf{t})\right\rceil,
\end{align}
where 
\begin{align}
p^k_{\scriptscriptstyle C}(c;\mathbf{t}) = \sum_{a\in\mathcal{A}} p^k_{\scriptscriptstyle C|A}(c|a;\mathbf{t}) p^{k-1}_{\scriptscriptstyle A|C}(a|c_{k-1};\mathbf{t})
\end{align}
is the marginalized rA2C over $\mathcal{A}$, $c_{k-1}$ is the $(k-1)$-th selected meaningful concept.
\end{corollary}
\begin{IEEEproof}
The proof is similar to the proof of Theorem \ref{prop:ShannonSNC} provided in Appendix \ref{appendix:proofofProp1}.
\end{IEEEproof}
In practice, it is hard to know the number of meaningful concepts $K$ that need to be selected for reliable System 2 SNC. However, as mentioned earlier, the greedy algorithm allows to select the meaningful concepts one-by-one, thereby allowing early stopping of communication when the reliability is guaranteed. Thus, the number of meaningful concepts of an entity in System 2 SNC is upper bounded by the number of extracted concepts from the same action in System 1 SNC. Such communication-efficiency of System 2 SNC is further corroborated by numerical results in the following section.
\section{Experimental Results}\label{sec:experiments}
{This section provides  experimental results to give more insights about System 1 and System 2 SNC. Especially, regarding System 1 SNC as a benchmark without contextual reasoning, we focus on demonstrating the effectiveness of System 2 SNC and contextual reasoning therein.} For the experiment, we fixed the number of actions and concepts in the world to $|\mathcal{A}| = 100$ and $|\mathcal{C}| = 100$. The singular A2Cs, each of which indicates the probability distribution of whether a concept is extracted from an action or not, are generated by the Dirichlet distribution with hyperparameter pair $(0.1,0.1)$. Both prior distributions of actions and concepts are uniform in the beginning of the communication, and might vary during communication in System 2 SNC. For System 1 SNC, we introduce a criterion to decide the  concepts extraction from an action, i.e., all concepts $c\in\mathcal{C}$ such that $p_{\scriptscriptstyle X_c|A}(\mathtt{TRUE}|a;t)\geq 0.9$ are extracted from a given intended action $a\in\mathcal{A}$. We assume a binary erasure channel (BEC) between the agents with erasure probability $p_e$ and the erased bits are retransmitted based on feedback (e.g., hybrid ARQ) to achieve the BEC capacity $1-p_e$ in probability. In the experiment, the communication reliability $\gamma$ is the ratio of the listener's correct inference about the speaker's intended action to the total communication rounds.


\begin{figure*}
\centering
\subfigure[$r = 20$]{\includegraphics[width=0.32\textwidth]{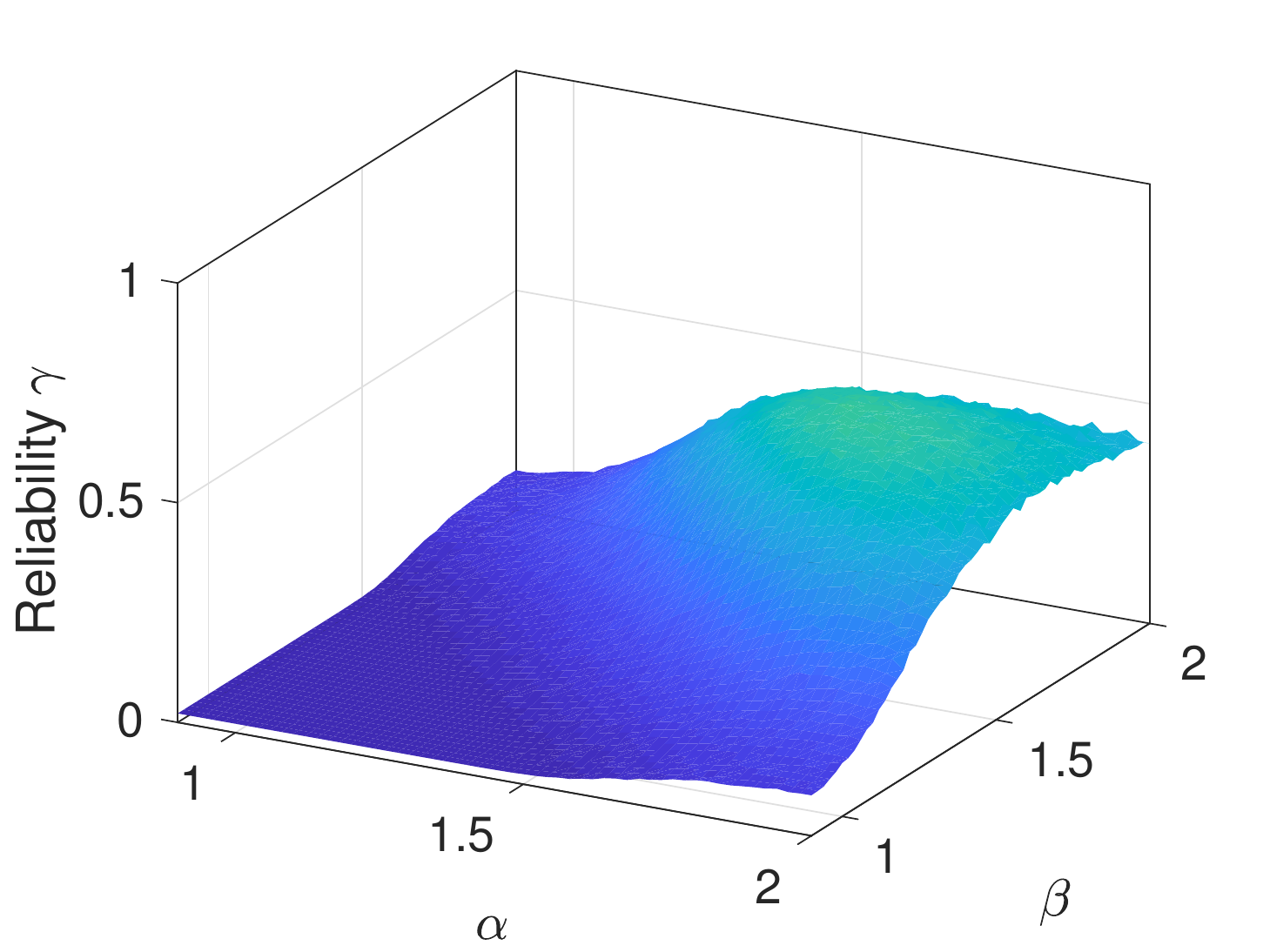}\label{fig:rSNCreliability_a}}
\subfigure[$r = 100$]{\includegraphics[width=0.32\textwidth]{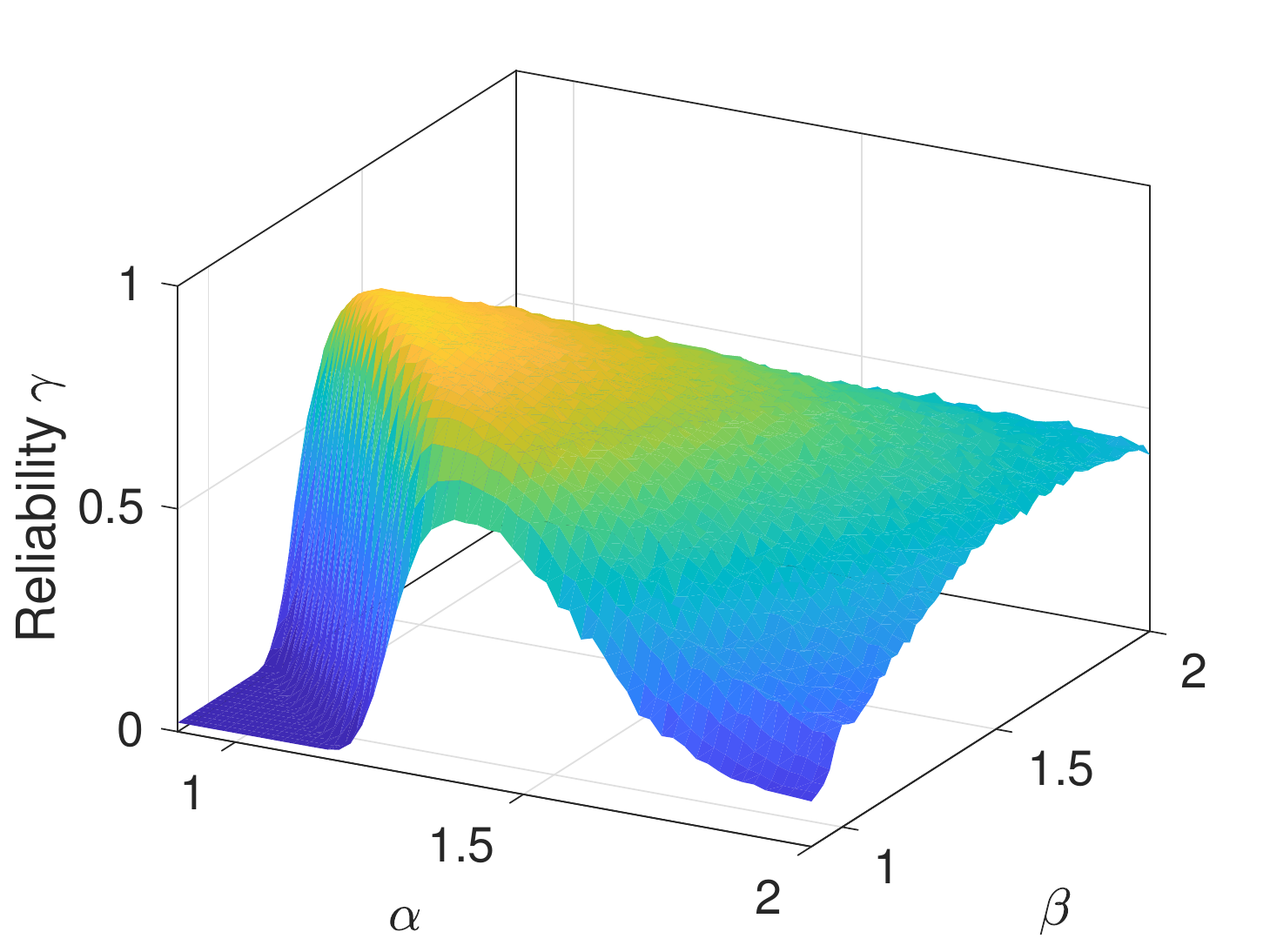}\label{fig:rSNCreliability_b}}
\subfigure[$r = 200$]{\includegraphics[width=0.32\textwidth]{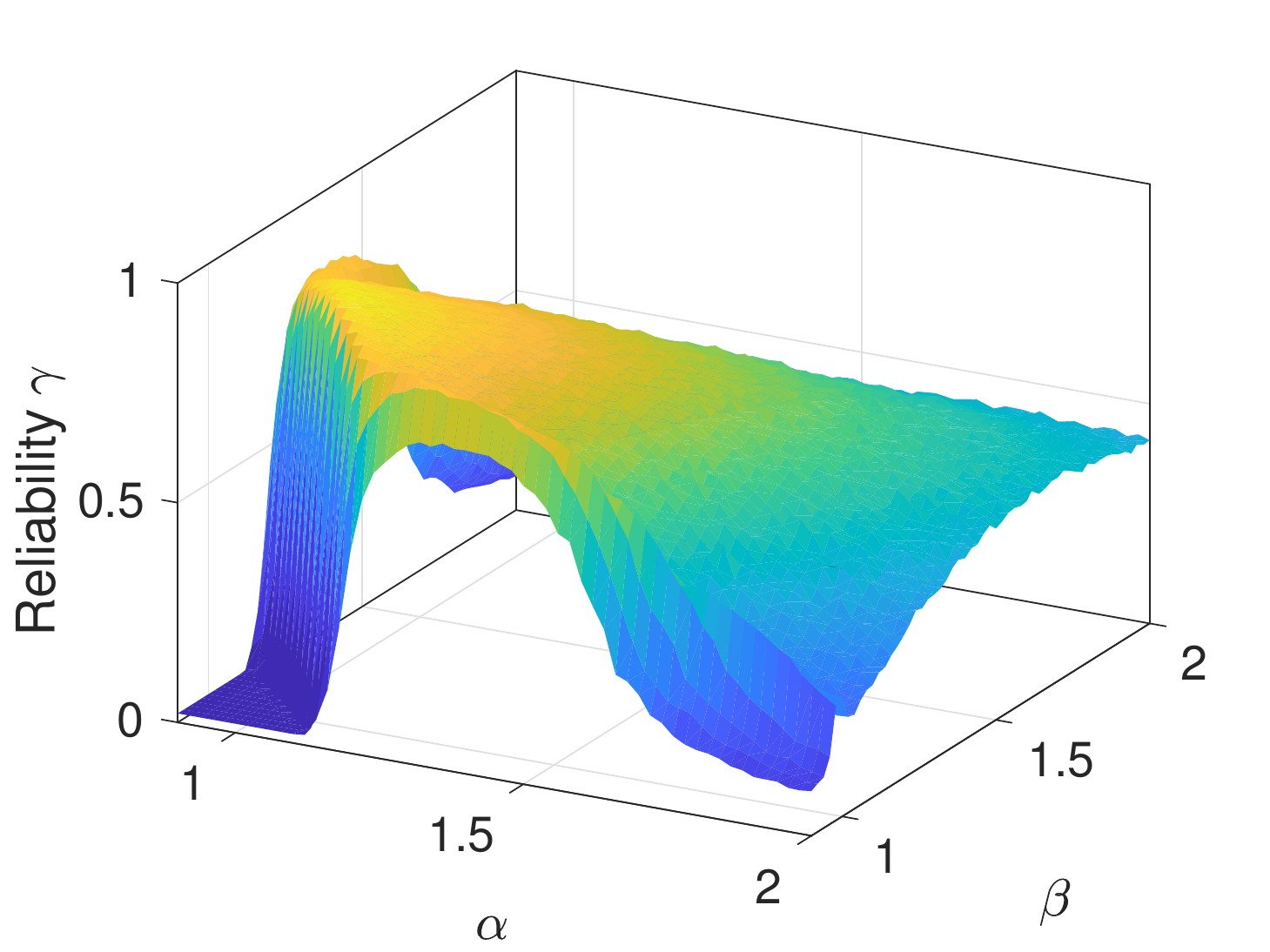}\label{fig:rSNCreliability_c}}
\caption{Reliability $\gamma$ with respect to parameters $\alpha$ and $\beta$ ranging $0.9$ to $2$ in System 2 SNC, for different self-SNC iteration depth $r = 10$, $20$ and $100$.}
\label{fig:rSNCreliability}
\end{figure*}

\begin{figure*}
\centering
\subfigure[$r=20$]{\includegraphics[width=0.32\textwidth]{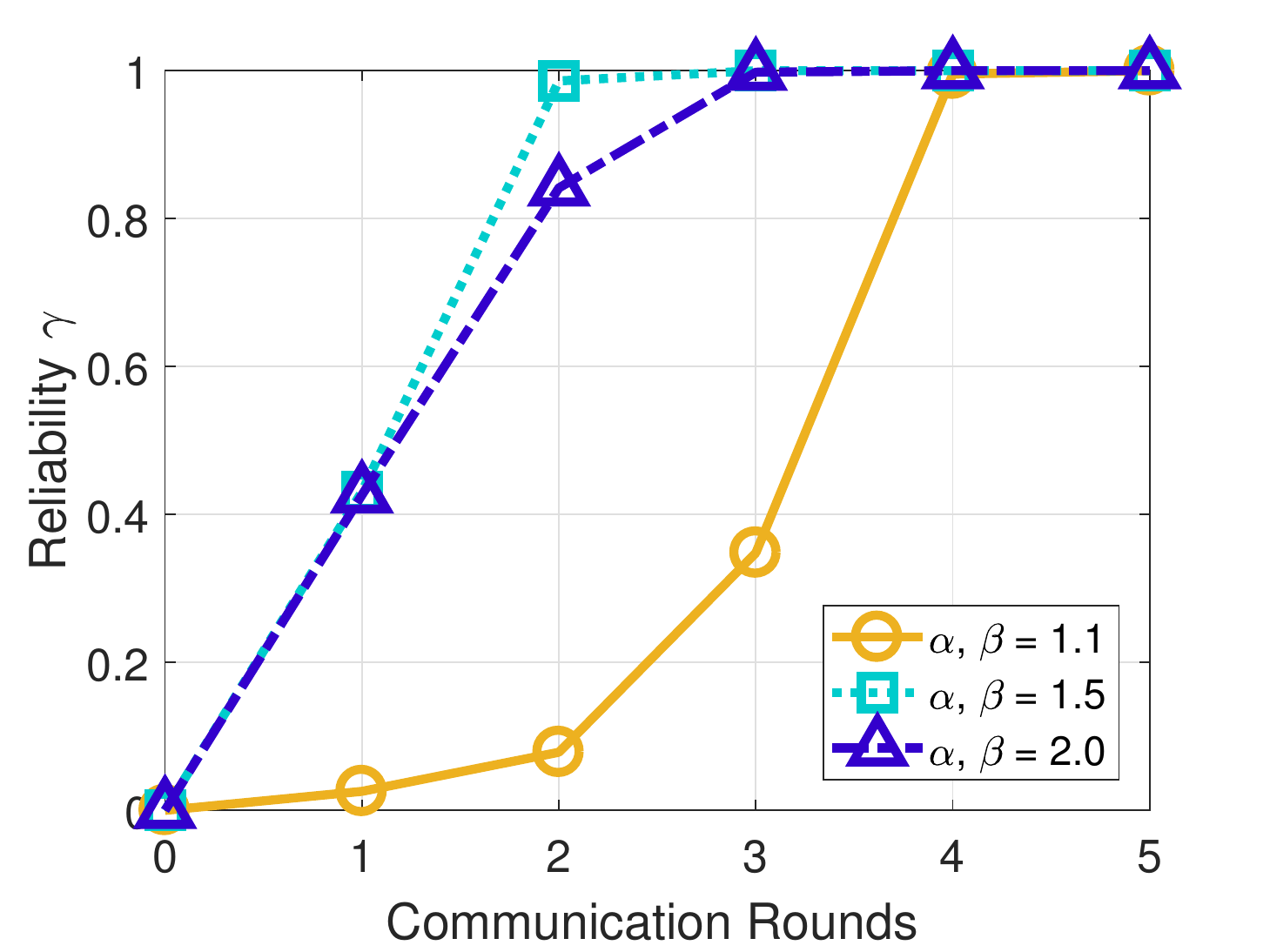}\label{fig:srSNCreliability_a}}
\subfigure[$r=100$]{\includegraphics[width=0.32\textwidth]{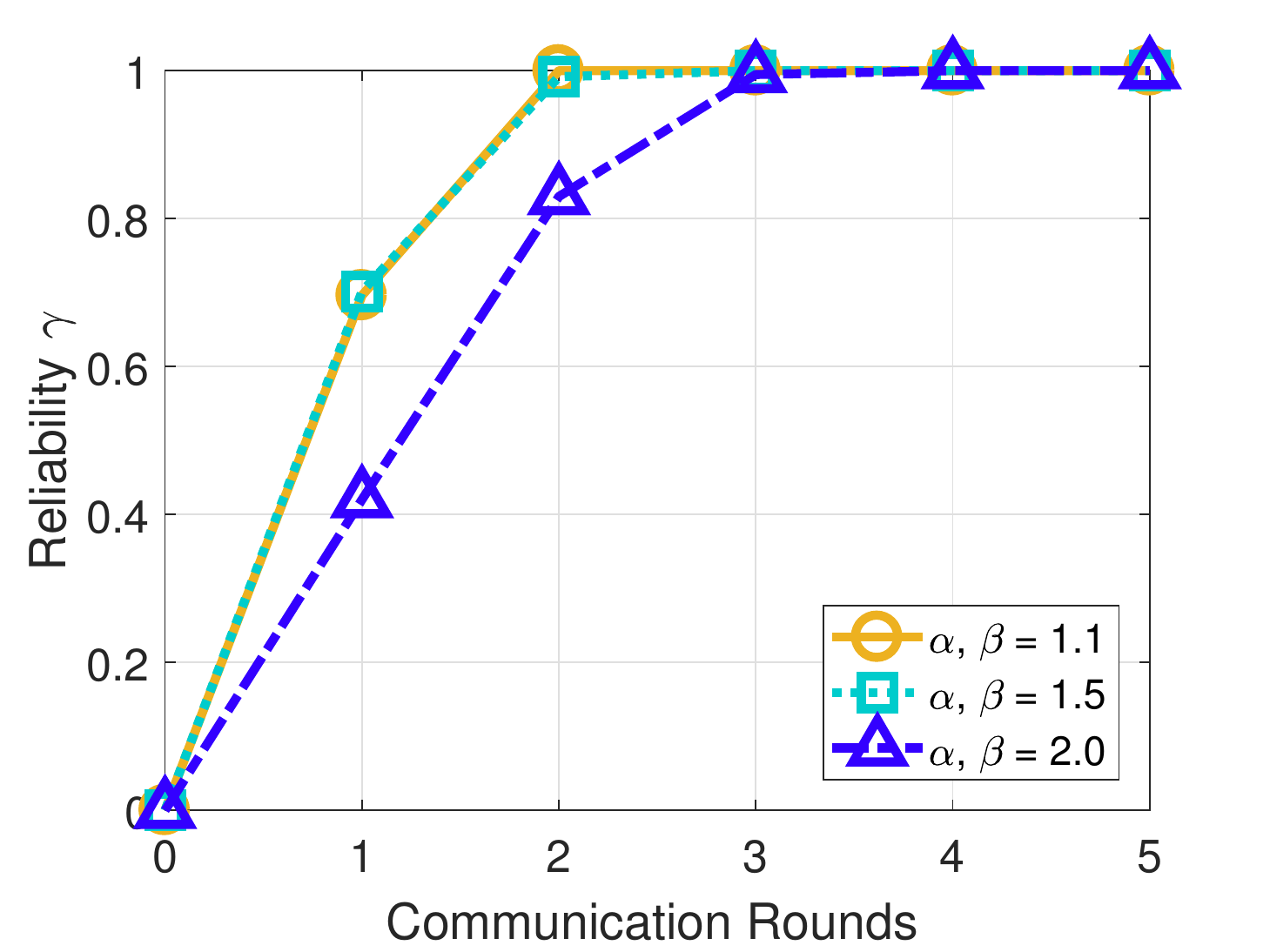}\label{fig:srSNCreliability_b}}
\subfigure[$r=200$]{\includegraphics[width=0.32\textwidth]{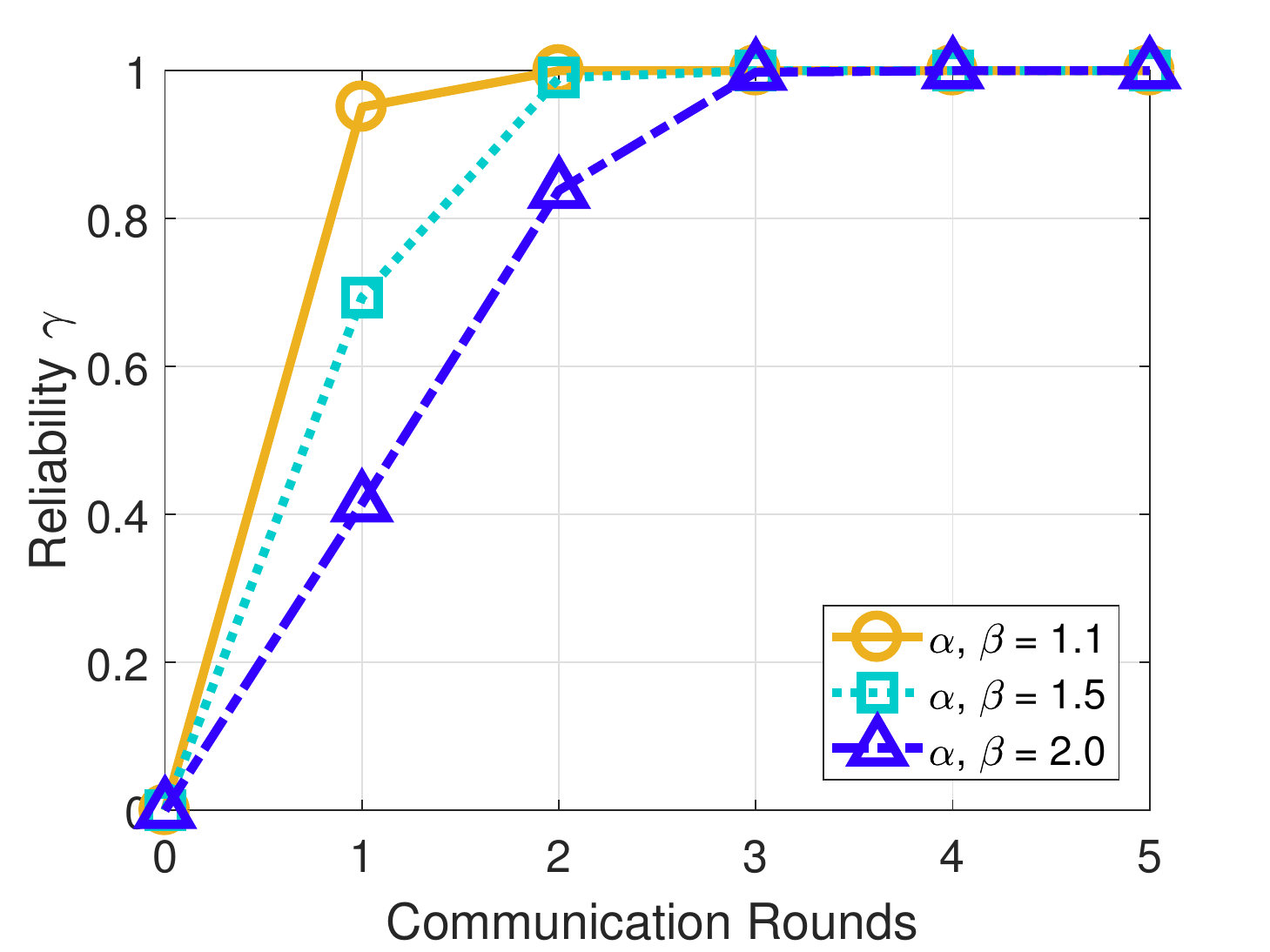}\label{fig:srSNCreliability_c}}
\caption{Reliability $\gamma$ versus communication rounds in System 2 SNC with different parameters $\alpha,\beta = 1.1$, $1.5$ and $2.0$, and different self-SNC iteration steps $r = 10$, $20$ and $200$ in each round.}
\label{fig:srSNCreliability}
\end{figure*}

\vspace{5pt}\noindent\textbf{Computation-Communication Trade-Off in System 2 SNC.}\quad
Fig. \ref{fig:rSNCreliability} shows the impact of self-SNC iteration steps on the reliability of System 2 SNC with a single meaningful concept. Here the iteration steps are directly related to the amount of computational resource that each agent use. As shown in Fig. \ref{fig:rSNCreliability}, for a larger number of iteration steps, i.e., with larger computational effort, the reliability of System 2 SNC is higher, especially when both parameters $\alpha$ and $\beta$ approach $1$. {This is related to the slow convergence of the self-SNC with $\alpha$ and $\beta$ close to 1 as shown in Fig. \ref{fig:iteration}. When $\alpha$ and $\beta$ are close to $1$ and the recursion is not converged, the reliability performance is poorer than the case when $\alpha$ and $\beta$ are close to $2$ and the recursion is converged. On the other hand, after convergence, the reliability performance of such a system becomes better when $\alpha$ and $\beta$ are closer to 1. Such computation-communication trade-off can be also found in Fig. \ref{fig:srSNCreliability}.} For instance, it can be easily found that with $\alpha, \beta = 1.1$, the communication reliability increases as the number of iteration steps increases. Moreover, the number of communicated meaningful concepts becomes smaller as the agents put more computational efforts in self-SNC.

\vspace{5pt}\noindent\textbf{Impact of $\alpha$ and $\beta$ on System 2 SNC.}\quad
As previously mentioned and shown in Fig. \ref{fig:iteration}, for the objective $\mathsf{G}$ in \eqref{eq:lossfunction} with larger $\alpha$ and $\beta$, the alternating iteration \eqref{eq:iteration_1}-\eqref{eq:iteration_4} approaches faster to the minimum. Such trend can be also seen in Figs. \ref{fig:rSNCreliability} and \ref{fig:srSNCreliability}, where with larger $\alpha, \beta$, the reliability of System 2 SNC with a single meaningful concept stays constant after $20$ iteration steps, while with smaller $\alpha, \beta$ it varies as the number of iteration steps increases until convergence. However, Fig. \ref{fig:rSNCreliability_a} shows that the reliability $\gamma$ does not exceed $0.5$ when $\alpha,\beta=2$, even after convergence. {On the other hand, when $\alpha$ and $\beta$ are  smaller, though it converges slowly, communication reliability is high after convergence. Specifically, under our experimental setting, the reliability $\gamma$ of the System 2 SNC approaches $1$ after 200 iteration steps even with a single meaningful concept, when $\alpha$ and $\beta$ are close to $1$ as shown in Fig. \ref{fig:rSNCreliability_c}. This provides the insight that faster convergence is not always better. Rather, slower and steadier reasoning provides better rationality in SNC. In other words, $\alpha$ and $\beta$ at each agent can be chosen based on the system reliability requirement and allowed computational resources regarding the computation-communication trade-off in System 2 SNC.} It is also worth noting  that at $\alpha, \beta = 1$, communication reliability is low since  minimizing \eqref{eq:lossfunctionKL} induces an uncertainty increase in both rA2C and rC2A.

\begin{figure*}
\centering
\subfigure[$p_e = 0$ (noiseless)]{\includegraphics[width=0.32\textwidth]{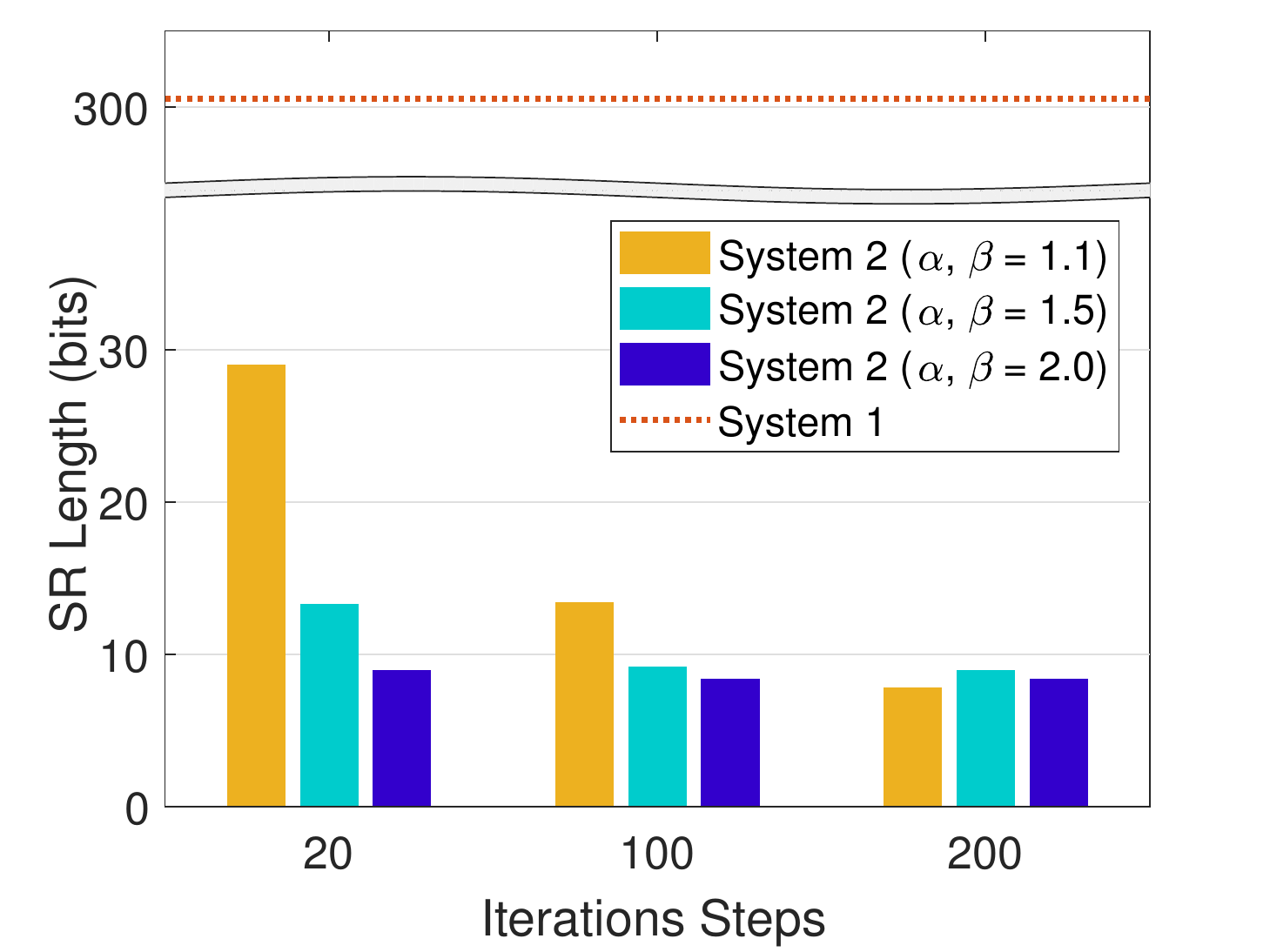}\label{fig:codelength_noiseless}}
\subfigure[$p_e = 0.1$]{\includegraphics[width=0.32\textwidth]{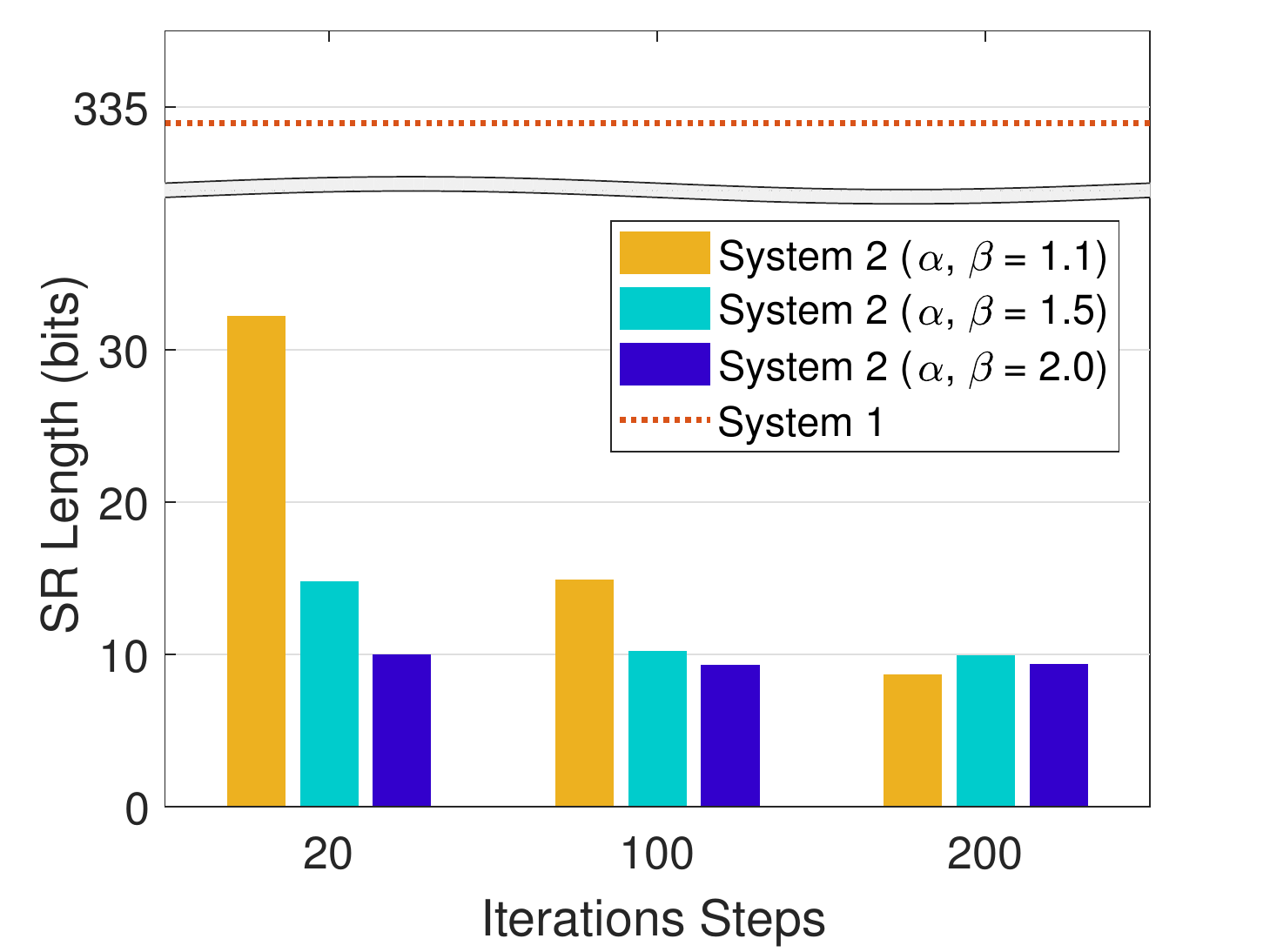}\label{fig:codelength_01}}
\subfigure[$p_e = 0.2$]{\includegraphics[width=0.32\textwidth]{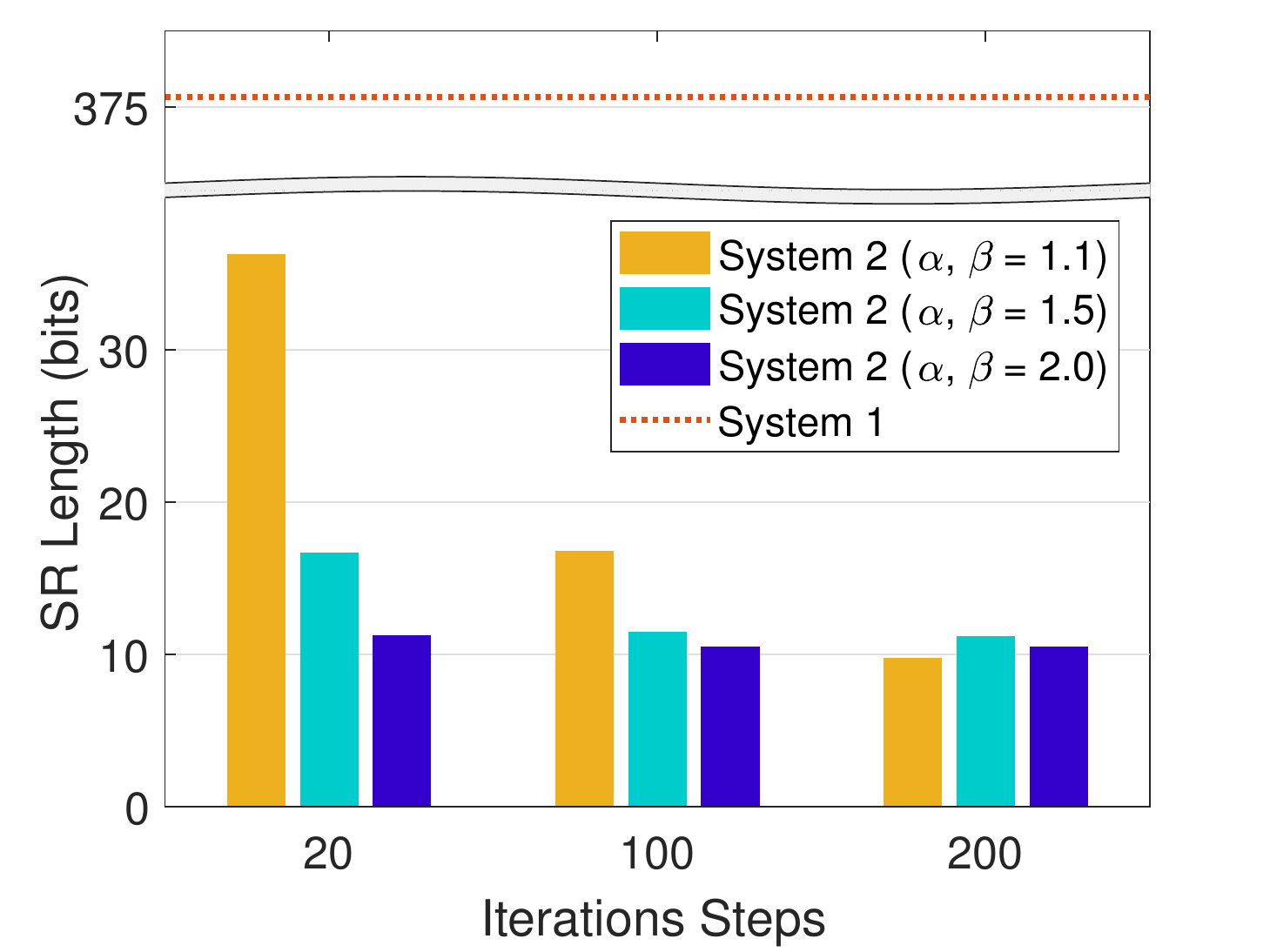}\label{fig:codelength_02}}
\caption{SR length achieving reliability $\gamma = 1$ in System 1 and System 2 SNC, under noiseless and noisy ($p_e = 0.1,0.2$ binary erasure) channels, for different self-SNC iteration steps $r = 10$, $20$ and $100$.}
\label{fig:codelength}
\end{figure*}

\vspace{5pt}\noindent\textbf{Reliability-Latency Trade-Off in System 2 SNC.}\quad
Fig. \ref{fig:srSNCreliability} illustrates the impact of the number of communication rounds on the reliability of System 2 SNC. Note that a meaningful concept is communicated in each round. The number of communication rounds therefore corresponds to the number of communicating meaningful concepts. Moreover, the larger number of communication rounds indicates higher communication latency. In Fig. \ref{fig:srSNCreliability_a}, each communication round runs for $20$ iteration steps of \eqref{eq:iteration_1}-\eqref{eq:iteration_4}, and thus the reliability of System 2 SNC when $\alpha,\beta = 1.1$ is poor compared to the other settings. However, the reliability $\gamma$ approaches to $1$ after $4$ communication rounds. As the iteration steps per communication round increases, the reliability of System 2 SNC with smaller $\alpha$ and $\beta$ enhances owing to their convergence as mentioned earlier. Furthermore, when $\alpha,\beta = 1.1$, the reliability $\gamma$ approaches $1$ within two communication rounds, which means that allocating more computational effort enhances the reliability-latency trade-off in System 2 SNC.

\vspace{5pt}\noindent\textbf{SR Length Comparison: System 1 SNC vs. System 2 SNC.}\quad 
The bit-length of the source-coded SR quantifies the size of SR in bits. In this regard, Fig. \ref{fig:codelength_noiseless} compares the expected bit-length of SRs in System 1 and System 2 SNC with noiseless channel between  communicating agents. Here, we consider cases in which SNC achieves a target reliability $\gamma = 1$. For System 1 SNC, the expected SR bit-length is related to the number of extracted concepts per action. Thus, in the experiment, the SR length exceeds $300$ bits in System 1 SNC. On the other hand, the expected length of coded SRs in System 2 SNC is significantly smaller that of System 1 SNC. One interesting aspect is that the average code length is always smaller when $\alpha,\beta = 2.0$ compared to the case $\alpha,\beta = 1.5$, even though it takes more communication rounds to reach reliability $\gamma = 1$ as shown in Fig. \ref{fig:srSNCreliability}. This is because, as also shown in Fig. \ref{fig:heatmap}, a larger value of $\alpha$ induces a reduction of uncertainty about what meaningful concepts should be chosen for System 2 SNC. Meanwhile, Figs. \ref{fig:codelength_01} and \ref{fig:codelength_02} show the total SR length including retransmissions under a noisy channel scenario when $p_e = 0.1$ and $p_e = 0.2$, respectively. Since  retransmissions with feedback can achieve the capacity of BEC, i.e., $1-p_e$, the results show that the SR length for reliability $\gamma = 1$ achieving System 1 and System 2 SNC is $\frac{1}{1-p_e}$ times longer than that with noiseless channel shown in Fig. \ref{fig:codelength_noiseless}.

\begin{figure}[!t]
	\centering
	\includegraphics[width=0.45\textwidth]{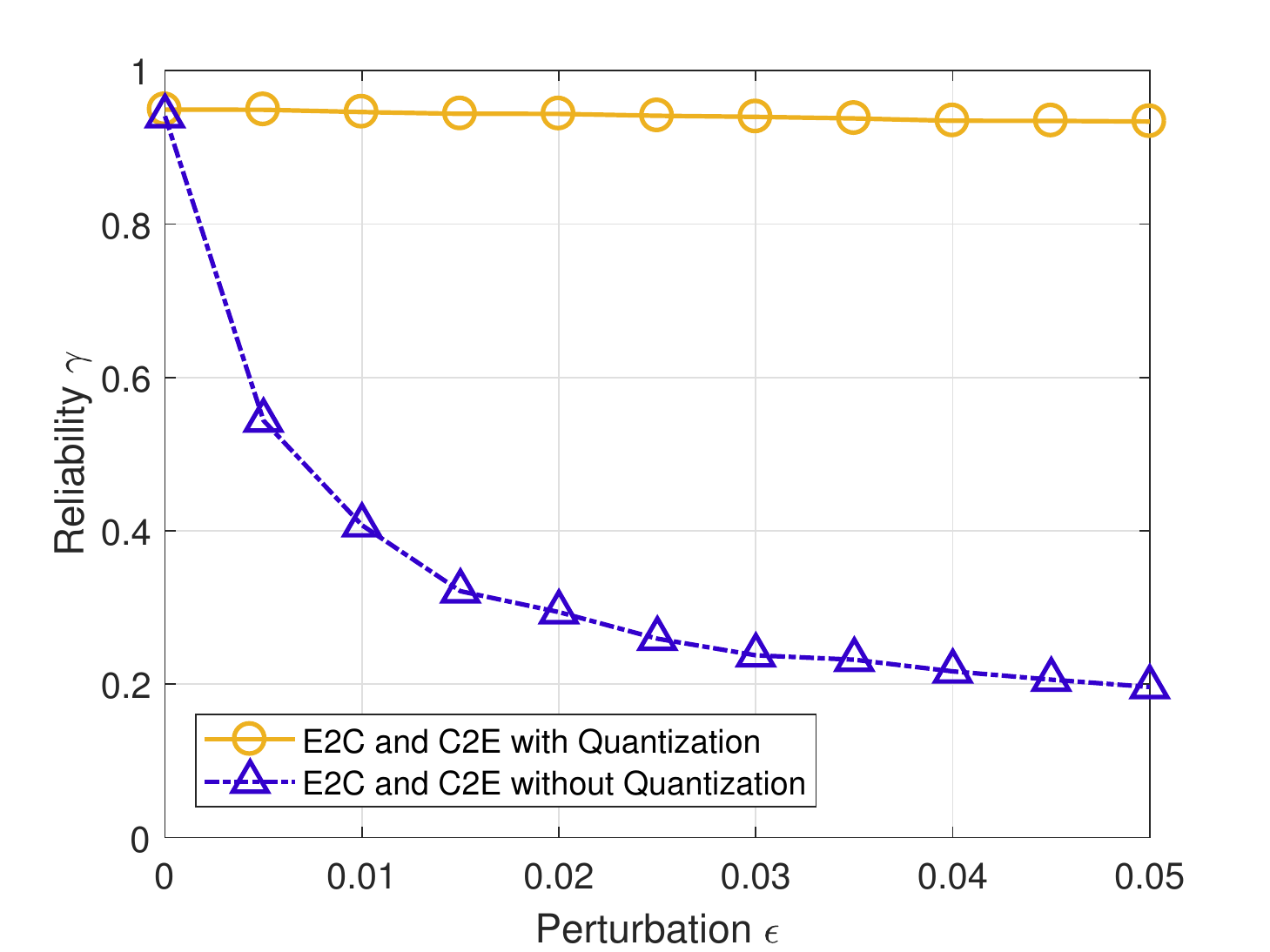}
	\caption{Reliability $\gamma$ of System 2 SNC with different initializations versus perturbed communication context.}
	\label{fig:perturbation}
\end{figure}

\vspace{5pt}\noindent\textbf{Robustness to Asynchronous Contextual Reasoning in System 2 SNC.}\quad Fig. \ref{fig:perturbation} illustrates the reliability of System 2 SNC with asynchronous self-SNC of both speaker and listener. Here, such asynchrony comes from speaker and listener having different A2C and C2A when initializing their self-SNC procedure. For a fixed speaker's A2C and listener's C2A, the experiment considers that there exists a random perturbation on the speaker's A2C known at the listener and listener's C2A known at the speaker. The amount of perturbation is chosen randomly and uniformly over $[-\epsilon,+\epsilon]$ for each component of A2C and C2A. When the perturbed A2C and C2A is directly used for initializing self-SNC at each agent, the reliability reduces significantly as $\epsilon$ increases as shown from the blue curve (\emph{A2C and C2A without Quantization}) in Fig. \ref{fig:perturbation}. {One way to ensure robustness against such perturbation is to quantize A2C and C2A before initializing the self-SNC at both speaker and listener. Here, we used a simple quantization method, i.e., rounding off the second decimal place of A2C and C2A entries.} The yellow curve (\emph{A2C and C2A with Quantization}) in Fig. \ref{fig:perturbation} shows the  reliability of System 2 SNC against perturbation when initializing the self-SNC with quantized A2C and C2A. Compared to the one without quantization,  System 2 SNC with self-SNC initialized with quantized A2C and C2A is more robust to perturbations.

\section{Conclusion}\label{sec:discussion}
In this article, we first proposed System 1 SNC model, which is a stochastic model for semantic communication among agents. Moreover, by infusing contextual reasoning into System 1 SNC, we developed a novel System 2 SNC model that extracts effective semantics for a given listener's communication context. Based on the proposed stochastic model, we numerically showed that System 2 SNC significantly reduces the SR bit-length with high reliability. Our proposed SNC framework and its stochastic model can be extended towards developing more effective SR, such as considering \emph{invariance} against nuisance factors or \emph{causal structure} of the agent tasks for more effective communication. Besides, the extension of the proposed theoretical framework can be carried towards more practical scenarios.
\appendices
\section{Proof of Theorem \ref{prop:ShannonSNC} and Corollary \ref{cor:3}} \label{appendix:proofofProp1}
A relative frequency distribution of extracting concept $c$ from the entities can be computed by 
\begin{align}\label{eq:frequency}
f_c = \frac{p_{\scriptscriptstyle X_c}(\mathtt{TRUE};t)}{\sum_{c\in\mathcal{C}}p_{\scriptscriptstyle X_{c}}(\mathtt{TRUE};t)},
\end{align}
for all $c\in\mathcal{C}$, and $\sum_{c\in\mathcal{C}} f_c = 1$, where where $p_{\scriptscriptstyle X_c}(x_c;t) = \sum_{a\in\mathcal{A}}p_{\scriptscriptstyle X_c|A}(x_c|a;t)p_{\scriptscriptstyle A}(a)$. Then, from Kraft's inequality, the expected length of coded symbols $s(1),s(2),\dots,s(|\mathcal{C}|)$ are lower bounded by the $d$-ary entropy with probabilities $f_1,f_2,\dots,f_{|\mathcal{C}|}$ as 
\begin{align}\label{eq:shannon_lower}
\sum_{c\in\mathcal{C}}f_c \,\ell_{\text{S$_1$},c} \geq -\sum_{c \in \mathcal{\mathcal{C}}} f_c \log_d f_c,
\end{align}
where $\ell_{\text{S$_1$},c}$ is the code length  for the symbol $s(c)$ for all $c\in\mathcal{C}$.
Thus, we have a lower bound
\begin{align}
\mathsf{L}_{\text{S$_1$}}(t) \label{eq:upperline}&= \sum_{c\in\mathcal{C}}p_{\scriptscriptstyle X_c}(\mathtt{TRUE};t)\ell_{\text{S$_1$},c}\\
\label{eq:middleline}&= \left(\sum_{c\in\mathcal{C}}p_{\scriptscriptstyle X_{c}}(\mathtt{TRUE};t) \right) \sum_{c\in\mathcal{C}}f_c\,\ell_{\text{S$_1$},c}\\
\label{eq:last_line_1}&\geq - \sum_{c\in\mathcal{C}}p_{\scriptscriptstyle X_{c}}(\mathtt{TRUE};t)\log_d f_c
\end{align}
where \eqref{eq:middleline} holds from \eqref{eq:frequency}, and the inequality \eqref{eq:last_line_1} holds from \eqref{eq:shannon_lower}. Consequently, by substituting \eqref{eq:frequency} into \eqref{eq:last_line_1}, we can obtain the lower bound \eqref{eq:lowerboundSNC}.

In the mean time, since $\log_d f_c$ is not always an integer, taking $\ell_{\text{S$_1$},c} = \lceil{-\log_d f_c}\rceil$ for all $c\in\mathcal{C}$ gives an upper bound
\begin{align}\label{eq:shannon_upper}
\sum_{c\in\mathcal{C}}f_c \,\ell_{\text{S$_1$},c} \leq \sum_{c \in \mathcal{\mathcal{C}}} f_c \lceil{-\log_d f_c}\rceil.
\end{align}
Thus, we also have an upper bound
\begin{align}
\label{eq:inequality_2}\text{\eqref{eq:middleline}} &\leq \sum_{c\in\mathcal{C}} p_{\scriptscriptstyle X_c}(\mathtt{TRUE};t) \sum_{c\in\mathcal{C}} f_c\, \lceil{-\log_d f_c}\rceil\\
\label{eq:last_line_2}&= \sum_{c\in\mathcal{C}} p_{\scriptscriptstyle X_{c}}(\mathtt{TRUE};t) \lceil{-\log_d f_c}\rceil,
\end{align}
where the inequality \eqref{eq:inequality_2} holds from \eqref{eq:shannon_upper}. Again, substituting \eqref{eq:frequency} into \eqref{eq:last_line_2} yields the upper bound \eqref{eq:upperboundSNC}, and this ends the proof.

Similarly, by replacing $f_c$ in \eqref{eq:last_line_1} and \eqref{eq:last_line_2} with $p^k_{\scriptscriptstyle C}(c;\mathbf{t}) = \sum_{a\in\mathcal{A}} p^k_{\scriptscriptstyle C|A}(c|a;\mathbf{t}) p^{k-1}_{\scriptscriptstyle A|C}(a|c_{k-1};\mathbf{t})$ and sum over $K$ communication rounds, we have the result in Corollary \ref{cor:3}.

\section{Proof of Theorem \ref{thm:theonly}}\label{appendix:proofofAM}
Next, we prove that \eqref{eq:iteration_1} to \eqref{eq:iteration_4} minimize the objective \eqref{eq:lossfunction} for $r\rightarrow \infty$, with given parameters $\alpha,\beta \geq 0$ and states $t$, $t'$ of two communicating agents. Let $\mathcal{P}$ be the set of all joint probability distributions of an action $A$ and concept $C$ parameterized by agent states $t$, $t'$. Note that the set is convex, since it is a set of probability distributions, which makes it possible to apply the alternating optimization of \eqref{eq:lossfunction} over it.

First, fix $\mathsf{S}^{\scriptscriptstyle [r-1]} \in \mathcal{P}$ and $\mathsf{L}^{\scriptscriptstyle [r-1]} \in \mathcal{P}$, thereby making \eqref{eq:lossfunction} a functional of $\mathsf{M} \in \mathcal{P}$, i.e., $\mathsf{G}(\mathsf{M})$. Note it can be easily shown that the functional is convex on $\mathcal{P}$. Since $\mathsf{M}$ is a probability distribution, a constraint $\sum_{(a, c) \in \mathcal{A}\times \mathcal{C}} \mathsf{M}(a,c;\mathbf{t}) = 1$ must be satisfied. Thus, we introduce a Lagrange multiplier $\gamma_{\scriptscriptstyle \mathsf{M}}$, and form a Lagrangian functional
\begin{align}\label{eq:lagrangian_1}
\mathsf{J}(\mathsf{M}) = \mathsf{G}(\mathsf{M}) - \sum_{(a,c)\in\mathcal{A} \times \mathcal{C}} \gamma_{\scriptscriptstyle \mathsf{M}} \mathsf{M}(a,c;\mathbf{t}).
\end{align}
Taking a derivative of \eqref{eq:lagrangian_1} with respect to $\mathsf{M}(a,c;\mathbf{t})$ gives
\begin{align}\label{eq:difflagrangian_1}
\frac{\partial \mathsf{J}(\mathsf{M}) }{\partial \mathsf{M}(a,c;\mathbf{t})} &= \left( \lambda \frac{\mathsf{S}^{\scriptscriptstyle [r-1]}(a,c;\mathbf{t})}{\mathsf{M}(a,c;\mathbf{t})} + (1-\lambda)\frac{\mathsf{L}^{\scriptscriptstyle [r-1]}(a,c;\mathbf{t})}{\mathsf{M}(a,c;\mathbf{t})} \right) - \gamma_{\scriptscriptstyle \mathsf{M}},
\end{align}
for all $(a,c) \in \mathcal{A}\times\mathcal{C}$. By equating the RHS of \eqref{eq:difflagrangian_1} to zero, we have
\begin{align}\label{eq:equatinglagrangian_1}
\mathsf{M}(a,c;\mathbf{r}) &= \frac{\lambda \mathsf{S}^{\scriptscriptstyle [r-1]}(a,c;\mathbf{t}) + (1 - \lambda) \mathsf{L}^{\scriptscriptstyle [r-1]}(a,c;\mathbf{t})}{\gamma_{\scriptscriptstyle \mathsf{M}}}.
\end{align}
Note that $\gamma_{\mathsf{M}} = \sum_{(a,c)\in\mathcal{A}\times\mathcal{C}} ( \lambda \mathsf{S}^{\scriptscriptstyle [r-1]}(a,c;\mathbf{t}) + (1 - \lambda) \mathsf{L}^{\scriptscriptstyle [r-1]}(a,c;\mathbf{t})) = 1$, since $\gamma_{\scriptscriptstyle \mathsf{M}}$ is the normalization constant in this case, and $\mathsf{S}$, $\mathsf{L}$ and $\mathsf{M}$ are defined on the same domain $\mathcal{A}\times\mathcal{C}$. Thus we have \eqref{eq:iteration_1}, that is $\mathsf{M}$ at recursion depth $r\geq 0$ before updating $\mathsf{S}$ and $\mathsf{L}$.

Now, fix $\mathsf{L}^{\scriptscriptstyle [r-1]}$ and $\mathsf{M}^{\scriptscriptstyle [r]}_{\scriptscriptstyle 1}$, making \eqref{eq:lossfunction} a functional of $\mathsf{S} \in \mathcal{P}$, i.e., $\mathsf{G}(\mathsf{S})$, where it can be also easily shown that $\mathsf{G}(\mathsf{S})$ is convex on $\mathcal{P}$. Considering a constraint $\sum_{(a, c) \in \mathcal{A}\times \mathcal{C}} \mathsf{S}(a,c;\mathbf{t}) = 1$, consider a Lagrangian functional
\begin{align}\label{eq:lagrangian_2}
\mathsf{J}(\mathsf{S}) = \mathsf{G}(\mathsf{S}) - \sum_{(a,c)\in\mathcal{A} \times \mathcal{C}} \gamma_{\scriptscriptstyle \mathsf{S}} \mathsf{S}(a,c;\mathbf{t}),
\end{align}
where $\gamma_{\scriptscriptstyle \mathsf{S}}$ is the Lagrangian multiplier. Taking the derivative of \eqref{eq:lagrangian_2} gives
\begin{align}\label{eq:difflagrangian_2}
\frac{\partial \mathsf{L}(\mathsf{S})}{\partial \mathsf{S}(a,c;\mathbf{t})} = -\lambda\left(\frac{\log \mathsf{S}(a,c;\mathbf{t}) + 1}{\alpha} - \log \mathsf{M}^{\scriptscriptstyle [r,1]}(a,c;\mathbf{t})  \right) - \gamma_{\scriptscriptstyle \mathsf{S}},
\end{align}
for all $(a,c) \in \mathcal{A}\times\mathcal{C}$. By equating the RHS of \eqref{eq:difflagrangian_2} to zero, we have
\begin{align}
    \log \mathsf{S}(a,c;\mathbf{t}) = \alpha \log \mathsf{M}^{\scriptscriptstyle [r]}_{\scriptscriptstyle 1}(a,c;\mathbf{t}) - \left(\frac{\alpha\gamma_{\scriptscriptstyle \mathsf{S}}}{\lambda} + 1\right).
\end{align}
Let $\frac{\alpha\gamma_{\scriptscriptstyle \mathsf{S}}}{\lambda} + 1 = \log Z_{\scriptscriptstyle \mathsf{S}}$, then we have
\begin{align}
\mathsf{S}(a,c;\mathbf{t}) = \frac{ \mathsf{M}^{\scriptscriptstyle [r]}_{\scriptscriptstyle 1}(a,c;\mathbf{t})^\alpha}{Z_{\scriptscriptstyle \mathsf{S}}}.
\end{align}
Here, $Z_{\scriptscriptstyle \mathsf{S}} = \sum_{(a,c)\in\mathcal{A}\times\mathcal{C}} \mathsf{M}^{\scriptscriptstyle [r]}_{\scriptscriptstyle 1}(a,c;\mathbf{t})^\alpha$ becomes a normalization constant. This yields \eqref{eq:iteration_2}.

The derivation of \eqref{eq:iteration_3} follows the same process of deriving \eqref{eq:iteration_1}, with the only difference  that $\mathsf{S}^{\scriptscriptstyle [r-1]}$ is updated to $\mathsf{S}^{\scriptscriptstyle [r]}$. Moreover, \eqref{eq:iteration_4} can be derived from a similar process of deriving \eqref{eq:iteration_2}, since $\mathsf{S}$ and $\mathsf{L}$ form a symmetry in \eqref{eq:lossfunction} (or nearly symmetric with different constants $\alpha$ and $\beta$). This ends the proof of the claim.

\section{Proof of Corollary \ref{cor:proofofPequalsQ}}\label{appendix:proofofPequalsQ}
After the convergence of \eqref{eq:iteration_1}-\eqref{eq:iteration_4}, 
\begin{align}\label{eq:convergeM}
\mathsf{M}^{\scriptscriptstyle[*]}(a,c;\mathbf{t}) = \lambda\mathsf{S}^{\scriptscriptstyle[*]}(a,c;\mathbf{t}) + (1-\lambda)\mathsf{L}^{\scriptscriptstyle[*]}(a,c;\mathbf{t})
\end{align}
holds for all $(a,c)\in\mathcal{A}\times\mathcal{C}$. By dividing both sides of \eqref{eq:convergeM} by $\mathsf{M}^{\scriptscriptstyle [*]}(a,c;\mathbf{t})$, we have
\begin{align}\label{eq:lineeq}
1 = \lambda x + (1-\lambda)y.
\end{align}
where $x = \frac{\mathsf{S}^{\scriptscriptstyle[*]}(a,c;\mathbf{t})}{\mathsf{M}^{\scriptscriptstyle[*]}(a,c;\mathbf{t})}$ and $y = \frac{\mathsf{L}^{\scriptscriptstyle[*]}(a,c;\mathbf{t})}{\mathsf{M}^{\scriptscriptstyle[*]}(a,c;\mathbf{t})}$. Note that from \eqref{eq:iteration_2} and \eqref{eq:iteration_4}, we have $\mathsf{S}^{\scriptscriptstyle [*]}(a,c;\mathbf{t}) = \frac{\mathsf{M}^{\scriptscriptstyle[*]}(a,c;\mathbf{t})^{\alpha}}{\sum\mathsf{M}^{\scriptscriptstyle[*]}(a,c;\mathbf{t})^{\alpha}}$ and $\mathsf{L}^{\scriptscriptstyle [*]}(a,c;\mathbf{t}) = \frac{\mathsf{M}^{\scriptscriptstyle[*]}(a,c;\mathbf{t})^{\beta}}{\sum\mathsf{M}^{\scriptscriptstyle[*]}(a,c;\mathbf{t})^{\beta}}$, respectively, after the convergence. Since we consider parameters $\alpha,\beta \geq 1$, it is always $x \leq 1$ and $y \leq 1$. By drawing \eqref{eq:lineeq} on the $x$-$y$ coordinate plane, it can be easily known that $(x,y) = (1,1)$ is the only point on the line that satisfies $x \leq 1$ and $y \leq 1$. Thus, $\mathsf{S}^{\scriptscriptstyle[*]}(a,c;\mathbf{t})=\mathsf{M}^{\scriptscriptstyle[*]}(a,c;\mathbf{t})$ and $\mathsf{L}^{\scriptscriptstyle[*]}(a,c;\mathbf{t})=\mathsf{M}^{\scriptscriptstyle[*]}(a,c;\mathbf{t})$, and thus $\mathsf{S}^{\scriptscriptstyle[*]}(a,c;\mathbf{t})=\mathsf{L}^{\scriptscriptstyle[*]}(a,c;\mathbf{t})$. This holds for all $(a,c)\in\mathcal{A}\times\mathcal{C}$, and it ends the proof.

\section{Proof of Theorem \ref{thm:reliability}}\label{appendix:proofofreliability}
To begin with, refer to the following lemma.
\begin{lemma}\label{lem:allnon-zero}
For any parameters $\alpha,\beta\geq1$, but not $(\alpha,\beta) = (1,1)$, and $0<\lambda<1$, for given converged $\mathsf{M}^{\scriptscriptstyle [*]}$, its all non-zero components, i.e., $\mathsf{M}^{\scriptscriptstyle [*]}(a,c;\mathbf{t})$ for some $(a,c)\in\mathcal{A}\times\mathcal{C}$ such that $\mathsf{M}^{\scriptscriptstyle [*]}(a,c;\mathbf{t}) \neq 0$, are equal to each other.
\end{lemma}
\begin{IEEEproof}
From Theorem \ref{thm:theonly} and Corollary \ref{cor:proofofPequalsQ}, $\mathsf{M}^{\scriptscriptstyle [*]}(a,c;\mathbf{t}) = \frac{\mathsf{M}^{\scriptscriptstyle [*]}(a,c;\mathbf{t})^\alpha}{\sum\mathsf{M}^{\scriptscriptstyle [*]}(a,c;\mathbf{t})^{\alpha}} = \frac{\mathsf{M}^{\scriptscriptstyle [*]}(a,c;\mathbf{t})^\beta}{\sum\mathsf{M}^{\scriptscriptstyle [*]}(a,c;\mathbf{t})^{\beta}}$. Since parameters $\alpha,\beta \geq 1$, but not $(\alpha,\beta) =(1,1)$, there always exists one is not equals to 1. Without loss of generality, let us say $\alpha > 1$. Then, for non-zero $\mathsf{M}^{\scriptscriptstyle [*]}(a,c;\mathbf{t})$, we have
\begin{align}\label{eq:sumMM}
\sum_{(a,c)\in\mathcal{A}\times\mathcal{C}}\mathsf{M}^{\scriptscriptstyle [*]}(a,c;\mathbf{t})^{\alpha} = \mathsf{M}^{\scriptscriptstyle [*]}(a,c;\mathbf{t})^{\alpha -1}.
\end{align}
For $\mathsf{M}^{\scriptscriptstyle [*]}(a,c;\mathbf{t}) = 1$, since $\sum_{(a,c)}\mathsf{M}^{\scriptscriptstyle [*]}(a,c;\mathbf{t}) = 1$, it is the only one that is non-zero. For $\mathsf{M}^{\scriptscriptstyle [*]}(a,c;\mathbf{t}) \neq 1$, since the LHS of \eqref{eq:sumMM} is fixed, we can conclude that $\mathsf{M}^{\scriptscriptstyle [*]}(a,c;\mathbf{t})$ for some $(a,c)\in\mathcal{A}\times\mathcal{C}$ such that $\mathsf{M}^{\scriptscriptstyle [*]}(a,c;\mathbf{t}) \neq 0$ are equal to each other.
\end{IEEEproof}
Note Lemma \ref{lem:allnon-zero} also applies for $\mathsf{S}^{\scriptscriptstyle [*]}$ and $\mathsf{L}^{\scriptscriptstyle [*]}$, from Corollary \ref{cor:proofofPequalsQ}. Now let $\mathsf{S}^{\scriptscriptstyle [*]}_k(a,c;\mathbf{t}) = p^k_{\scriptscriptstyle C|A}(c|a;\mathbf{t})p^{k-1}_{\scriptscriptstyle A|C}(a|c_{k-1};\mathbf{t})$ and $\mathsf{L}^{\scriptscriptstyle [*]}_k(a,c;\mathbf{t}) = p^{k}_{\scriptscriptstyle A|C}(a|c;\mathbf{t})p_{\scriptscriptstyle C}(c)$ for all $a\in\mathcal{A}$ and $c\in\mathcal{C}$ be the stationary individual CCs of the speaker and listener at $i$-th communication round of System 2 SNC, where $c_{k-1}$ is the communicated meaningful concept at $k-1$-th communication round. Then from Corollary 1, by equating $\mathsf{S}^{\scriptscriptstyle [*]}_k(a,c;\mathbf{t})$ and $\mathsf{L}^{\scriptscriptstyle [*]}_k(a,c;\mathbf{t})$, and dividing both sides with $p^{k}_{\scriptscriptstyle C}(c)$ defined over $\mathcal{C}_{k-1}$, such that $\mathcal{C}_k = \mathcal{C}_{k-1}\backslash c_{k}$ for $k\geq 1$ and $\mathcal{C}_0 = \mathcal{C}$, we have
\begin{align}
p^k_{\scriptscriptstyle A|C}(a|c;\mathbf{t}) = \frac{p^k_{\scriptscriptstyle C|E}(c|a;\mathbf{t})}{p^{k}_{\scriptscriptstyle C}(c)}p^{k-1}_{\scriptscriptstyle A|C}(a|c_{k-1};\mathbf{t}).    
\end{align}
Since $p^{k}_{\scriptscriptstyle C|A}(c|a;\mathbf{t})$ is defined over a reduced set $\mathcal{C}_{k-1}$ and Lemma \ref{lem:allnon-zero} suggests that all non-zero components are the same, we know that $p^{k}_{\scriptscriptstyle C|A}(c|a;\mathbf{t})\geq p^{k}_{\scriptscriptstyle C}(c)$ for all $c\in\mathcal{C}$, since $p^{k}_{\scriptscriptstyle C}(c)$ is uniform by definition. Therefore, $p^{k}_{\scriptscriptstyle A|C}(a|c;\mathbf{t}) \geq p^{k-1}_{\scriptscriptstyle A|C}(a|c_{k-1};\mathbf{t})$ for all $a\in\mathcal{A}$ and this ends the proof.

\bibliographystyle{IEEEtran}
\bibliography{IEEEabrv,Pragmatic}
\end{document}